\documentclass[preprint, 12pt,review,3p]{elsarticle}

\usepackage{amsmath}
\usepackage{amssymb}
\usepackage{gensymb}
\usepackage{graphicx}
\usepackage{lmodern}
\usepackage{caption}
\usepackage[font=footnotesize]{subcaption}
\usepackage[shortlabels]{enumitem}
\usepackage[]{units}
\usepackage{setspace}
\usepackage{booktabs}
\usepackage{tabu}
\usepackage{appendix}
\usepackage[pdftex,colorlinks,bookmarksopen,bookmarksnumbered,citecolor=red,urlcolor=red]{hyperref}

\usepackage{enumitem} 							
\usepackage{gensymb} 								
\usepackage{listings} 							

\usepackage[labelfont=bf]{caption} 	
\usepackage{subcaption}									
\usepackage{epstopdf}								

\usepackage{float}									

\usepackage{bm}
\usepackage{cleveref}
\usepackage{wrapfig}
\usepackage{multirow}
\usepackage{physics}
\usepackage{xfrac}
\usepackage{booktabs}
\usepackage{tikz}
\usetikzlibrary{shapes.geometric, arrows}
\tikzstyle{startstop} = [rectangle, rounded corners, minimum width=3cm, minimum height=1cm,text centered, draw=black, fill=white!20]
\tikzstyle{io} = [trapezium, trapezium left angle=70, trapezium right angle=110, minimum width=3cm, minimum height=1cm, text centered, draw=black, fill=white!20]
\tikzstyle{process} = [rectangle, minimum width=3cm, minimum height=1cm, align=center, draw=black, fill=white!20]
\tikzstyle{decision} = [diamond, minimum width=3cm, minimum height=1cm, text centered, text width=3cm, draw=black, fill=white!20]
\tikzstyle{arrow} = [thick,->,>=stealth]
\let\OldTextregistered\textregistered
\renewcommand{\textregistered}{\OldTextregistered\xspace}%
\newcommand\figref{Figure~\ref}

\biboptions{sort&compress}

\journal{Composites Part A: Applied Science and Manufacturing}

\begin{document}
\sloppy

\begin{frontmatter}

\title{Overcoming the cohesive zone limit in the modelling of composites delamination with TUBA cohesive elements}

\author[TUD]{Giorgio~Tosti~Balducci}
\author[TUD]{Boyang~Chen\corref{Chen}\fnref{Chen}}
\ead{b.chen-2@tudelft.nl}
\cortext[Chen]{Corresponding author}

\address[TUD]{Department of Aerospace Structures and Materials, Faculty of Aerospace Engineering, Delft University of Technology. Kluyverweg 1, 2629 HS Delft, The Netherlands}

\begin{abstract}
The wide adoption of composite structures in the aerospace industry requires reliable numerical methods to account for the effects of various damage mechanisms, including delamination. Cohesive elements are a versatile and physically representative way of modelling delamination. However, using their standard form which conforms to solid substrate elements, multiple elements are required in the narrow cohesive zone, thereby requiring an excessively fine mesh and hindering the applicability in practical scenarios. The present work focuses on the implementation and testing of triangular thin plate substrate elements and compatible cohesive elements, which satisfy $C^1$-continuity in the domain. The improved regularity meets the continuity requirement coming from the Kirchhoff Plate Theory and the triangular shape allows for conformity to complex geometries. The overall model is validated for mode I delamination, the case with the smallest cohesive zone. Very accurate predictions of the limit load and crack propagation phase are achieved, using elements as large as 11 times the cohesive zone.

\end{abstract}


\end{frontmatter}
%
%
\section{Introduction}
\label{sec:Int}
\begin{figure}
	\centering
	\includegraphics[width=\textwidth]{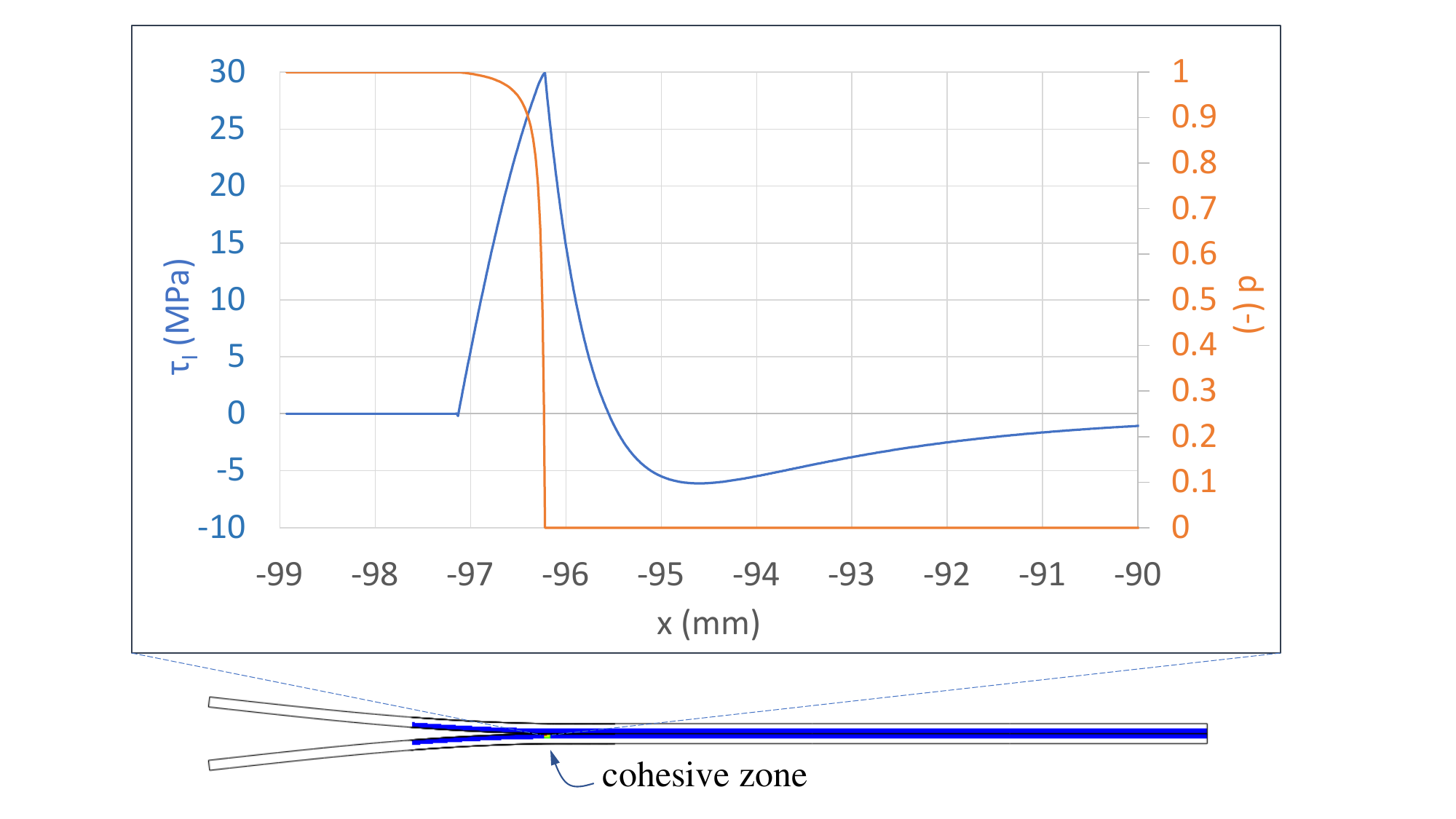}
	\caption{Stress and damage profiles near the crack tip of a 2D DCB specimen, obtained with Abaqus cohesive contact analysis. The element length is 0.0125 mm in the propagation region.} 
	\label{fig:stress_damage_2d}
\end{figure}

Composite materials constitute the current paradigm for efficient structural solutions. Their anisotropic architecture can in principle be designed to achieve an optimal distribution of strength and stiffness properties, while ensuring low weight of the final product. The laminated layout of composite structures allows for versatility in tuning the mechanical properties, but inevitably introduces weak interfaces between adjacent plies, known as the resin rich regions. Delamination would form at these interfaces upon a critical level of loading and could eventually lead to structural failure. 
The ability of Cohesive Elements (CEs) to predict crack onset and propagation for both ductile and brittle materials has made them an appealing tool in delamination analysis. They can be conveniently placed along the interfaces to capture the potential delamination. However, its widespread adoption in the industrial practice is hindered by a stringent mesh density requirement. A minimum number of elements, ranging in literature from $3$ to $10$ \cite{Harper2008CohesiveDelamination,Camanho2002,Camanho2002-NASA,Yang2005CohesiveComposites,Moes2002ExtendedGrowth} is in fact needed in the so-called  {cohesive zone}, which generally extends ahead of the crack tip for a few millimeters in composites delamination. There are many work in the literature on estimating the cohesive zone length so as to set appropriate mesh density for the analysis a priori, its accurate prediction however remains illusive \cite{Turon2007AnalyticalMaterials,Harper2008CohesiveDelamination,Soto2016CohesiveDelamination}. The most stringent case is the Mode I delamination, where the cohesive zone can be less than 1 mm long (Figure~\ref{fig:stress_damage_2d}) and CEs less than 0.5 mm would be required. This cohesive zone limit on mesh density arises from the inability of standard CEs to reproduce the steep stress gradients in the cohesive zone, except when extremely small elements are used.

\begin{figure}
	\centering
 \includegraphics[width=\textwidth]{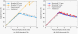}
	\caption{DCB load-displacement curves obtained with structural CEs \cite{RussoChen2019Jarticle}. Standard CE results, analytical solution and experimental data are also reported for comparison.} 
	\label{fig:DCB_raf_3rdOrd_Ld}
\end{figure}

Many works have been done in the past on tackling the cohesive zone limit on the mesh density of finite element models with CEs \cite{Turon2007AnModels,Guiamatsia2009Decohesion,Guiamatsia2011AFramework,Samimi2009enriched,Samimi2011self,Samimi2011three,Yang2010AnAnalyses,Do2013ImprovedElements,Alvarez2014ModeElements,Vandermeer2012Levelset,Latifi2017Thicklevelset,LU2018AdaptiveFNM,Irzal2014AnProblems,RussoChen2019Jarticle,daniel2023efficient,mukhopadhyay2023accurate}, an exhaustive review is not in the scope of this paper. The damage variables and stresses in a CE are determined by its opening vector. A finite element model of a composite laminate with delamination must be able to reproduce accurately the deformation of the sublaminates on both sides of the CE, if the delamination onset and progression are to be well predicted. As composite laminates are predominantly thin plates, the Kirchhoff Plate Theory can provide an accurate description of their deformation. The earlier work by Russo and Chen \cite{RussoChen2019Jarticle} has shown promising results in 2D, where the Euler-Bernoulli beam elements have been used to model the plies and a compatible CE, named structural CE, has been developed to model delamination. This contrasts with the kinematics of the standard CE, where no structural mechanics of the slender substrates have been taken into account. Their work has shown that the structural CE approach can indeed overcome the cohesive zone limit in 2D delamination analysis, allowing elements much larger than the cohesive zone to be used in the mesh, as shown in Figure~\ref{fig:DCB_raf_3rdOrd_Ld}. 

The aim of the present research is to extend this earlier work to 3D, where we will implement a suitable $C^1$-continuous thin plate element to model the plies, and develop a compatible CE to model the interfaces. The two elements will have triangular support, since this can conform to complex geometries, object of possible future studies. The objective of the present work is to assess the accuracy and computational performances of the proposed modelling method on the case with the most stringent mesh requirement due cohesive zone limit - the Mode I delamination. The Double Cantilever Beam bending (DCB) problem will be used as the benchmark.


The contents of this paper are organized as follows. Section \ref{sec:PS} presents the proposed $C^{1}$ triangular plate element for substrates and develops the associated CE for delamination. Section \ref{sec:VL} validates the proposed method on the DCB delamination problem. Finally, Section \ref{sec:CFW} summarizes the work performed, draws conclusions based on the results obtained and discusses possible future developments.


\section{Proposed Method}
\label{sec:PS}
\par The separating arms of a delaminating composite laminate are usually slender elements that, under the applied load, deform either as Euler-Bernoulli beams in two dimensions or as Kirchhoff plate/shells in a 3D space. These structural theories require the deflection to be $C^1$-continuous over the domain \cite{Zienkiewicz2013TheFundamentals}, thereby making its formulation more difficult than those of the standard $C^0$-continuous solid elements. In the rest of the section, a triangular thin-plate element, known as TUBA3 or Bell's triangle \cite{Bell1969AElement}, which satisfies the $C^1$-continuity requirement, is presented in detail. Afterwards, the TUBA3-compatible CE, hereby named TUBA3-CE, is derived. Both the TUBA3 plate element and the compatible TUBA3-CE were implemented as Abaqus user-element subroutines \cite{Abaqus2009}, written in the Fortran 95 programming language.

\subsection{The $C^1$ TUBA3 plate element}
\label{sec:tuba3_plate}
\par A famous class of triangular plate bending elements is the so-called TUBA$n$ family, first proposed by Argyris \textit{et al.} \cite{Argyris1968TheMethod}. Each TUBA$n$ element has $n$ nodes and presents the following features
\begin{enumerate}
    \item The out-of-plane displacement $w$ and its gradient are continuous inside the element and at its boundary. In other words, all the TUBA$n$ elements are of class $C^1$.
    \item Among the Degrees of Freedom (DOFs) of these elements are the bending and twisting curvatures at the corner nodes
    \begin{equation}
        \bm{k_i}=
        \begin{bmatrix}
            \pdv[2]{w}{x}\vert_i & \pdv[2]{w}{y}\vert_i & 2\pdv{w}{x}{y}\vert_i
        \end{bmatrix}^T
        \mbox{\hspace{7mm} $i=$ 1,2,3 }
    \end{equation}
\end{enumerate}
Having the curvatures defined at the nodes, it follows that stresses and strains are continuous at the nodes, since
\begin{align}
    \bm{\varepsilon} &= -z\bm{k}\\
    \bm{\sigma} &= \bm{C}\bm{\varepsilon}
\end{align}

\subsubsection{TUBA3: DOFs and shape functions}
\label{ssec:tuba3_dofs_sfuns}
\begin{figure}
	\centering
	\includegraphics[width=0.8\textwidth]{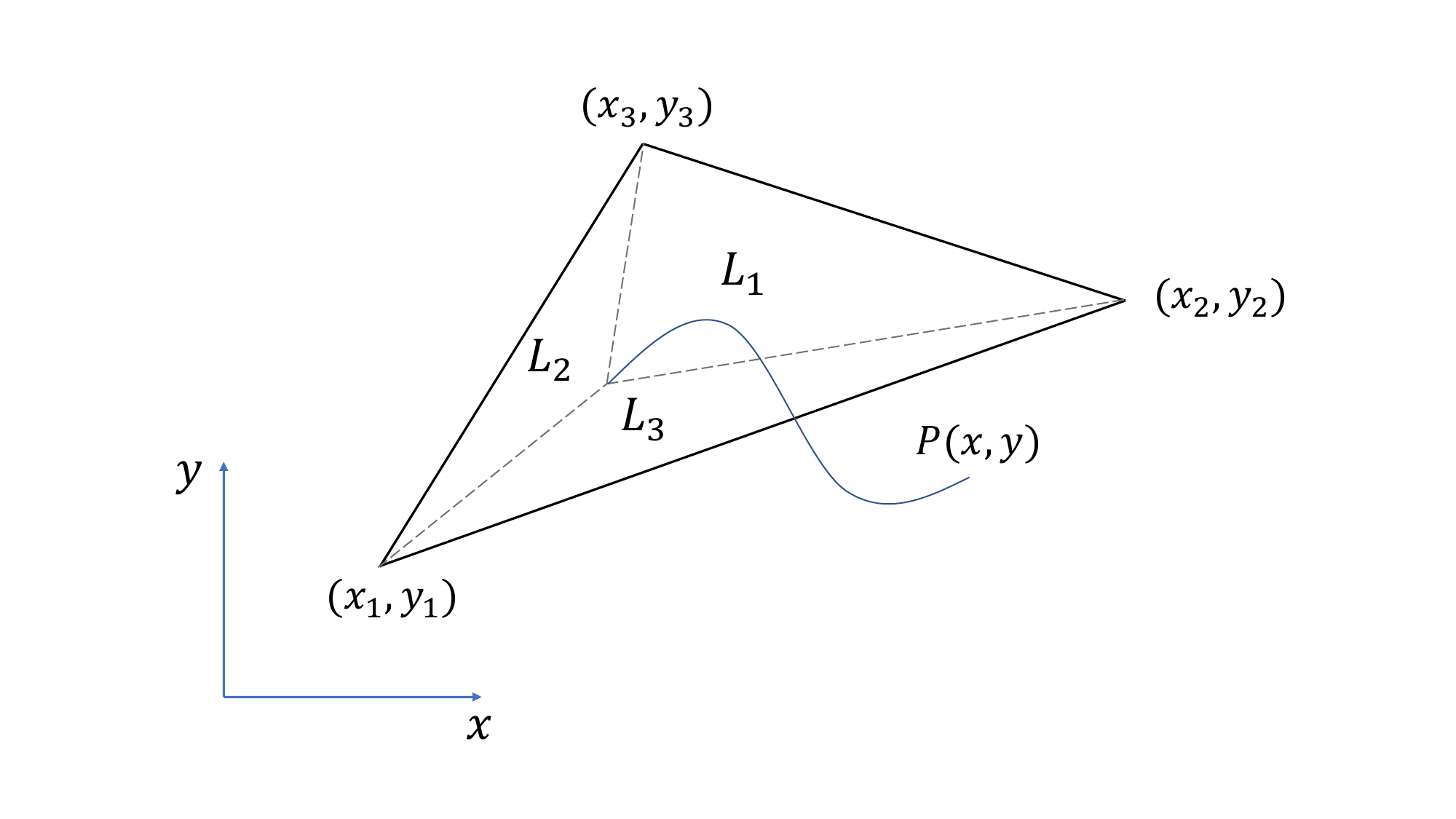}
	\caption{Area coordinates of a point in the generic triangle.} 
	\label{fig:ps_area_coord}
\end{figure}
The TUBA$n$ element chosen for this work is the three-noded TUBA3, also known as Bell's triangle \cite{Bell1969AElement}. This element is derived from the renowned TUBA6 - the Argyris triangle. The interpolation of the geometry and displacement field for a triangular element is often done in terms of area coordinates (\figref{fig:ps_area_coord}). These relate to the Cartesian ones in the following way
\begin{equation}
    \begin{cases}
        x=&L_1x_1+L_2x_2+L_3x_3\\
        y=&L_1y_1+L_2y_2+L_3y_3\\
        1=&L_1+L_2+L_3
    \end{cases}
    \label{eq:x_from_area_coords}
\end{equation}
where $(x_1,y_1)$, $(x_2,y_2)$ and $(x_3,y_3)$ are the triangle's corners in the Cartesian reference. The mapping in \Cref{eq:x_from_area_coords} can be inverted to obtain the area coordinates as
\begin{equation}
    \begin{cases}
        L_1=&\frac{1}{2A}(a_1+b_1x+c_1y)\\
        L_2=&\frac{1}{2A}(a_2+b_2x+c_2y)\\
        L_3=&1-L_1-L_2
    \end{cases}
    \label{eq:area_coords_from_x}
\end{equation}
The coefficients $a_i$, $b_i$ and $c_i$ read 
\begin{equation}
    \begin{aligned}
        a_i=&x_jy_k-x_ky_j\\
        b_i=&y_j-y_k\\
        c_i=&x_k-x_j
    \end{aligned}
    \label{eq:abc}
\end{equation}
where the indices $i$,$j$ and $k$ are cyclic permutations of 1, 2 and 3. The term $A$ in \Cref{eq:area_coords_from_x} is the triangle's area, computed from the corners coordinates as
\begin{equation}
    A=\frac{1}{2}\abs{\text{det}
    \begin{bmatrix}
        1&x_1&y_1\\
        1&x_2&y_2\\
        1&x_3&y_3
    \end{bmatrix}}
    \label{eq:triangle_area}
\end{equation}
\par
\begin{figure}
    \begin{minipage}[b]{0.5\linewidth}
        \centering
        \includegraphics[width=\textwidth]{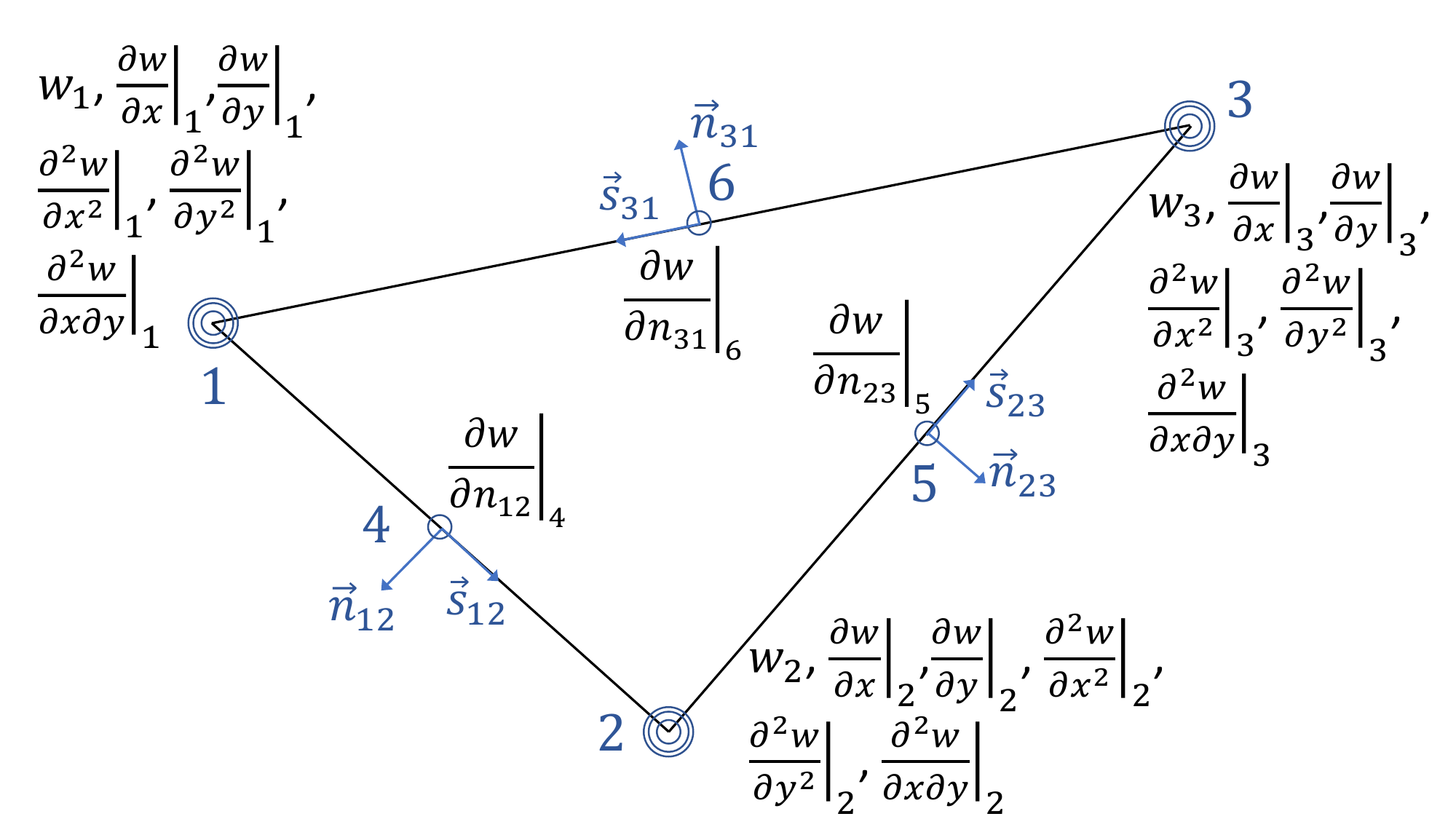}
        \caption{Argyris triangle (TUBA6).}
        \label{fig:ps_tuba6}
    \end{minipage}
    \hspace{0.5cm}
    \begin{minipage}[b]{0.5\linewidth}
        \centering
        \includegraphics[width=\textwidth]{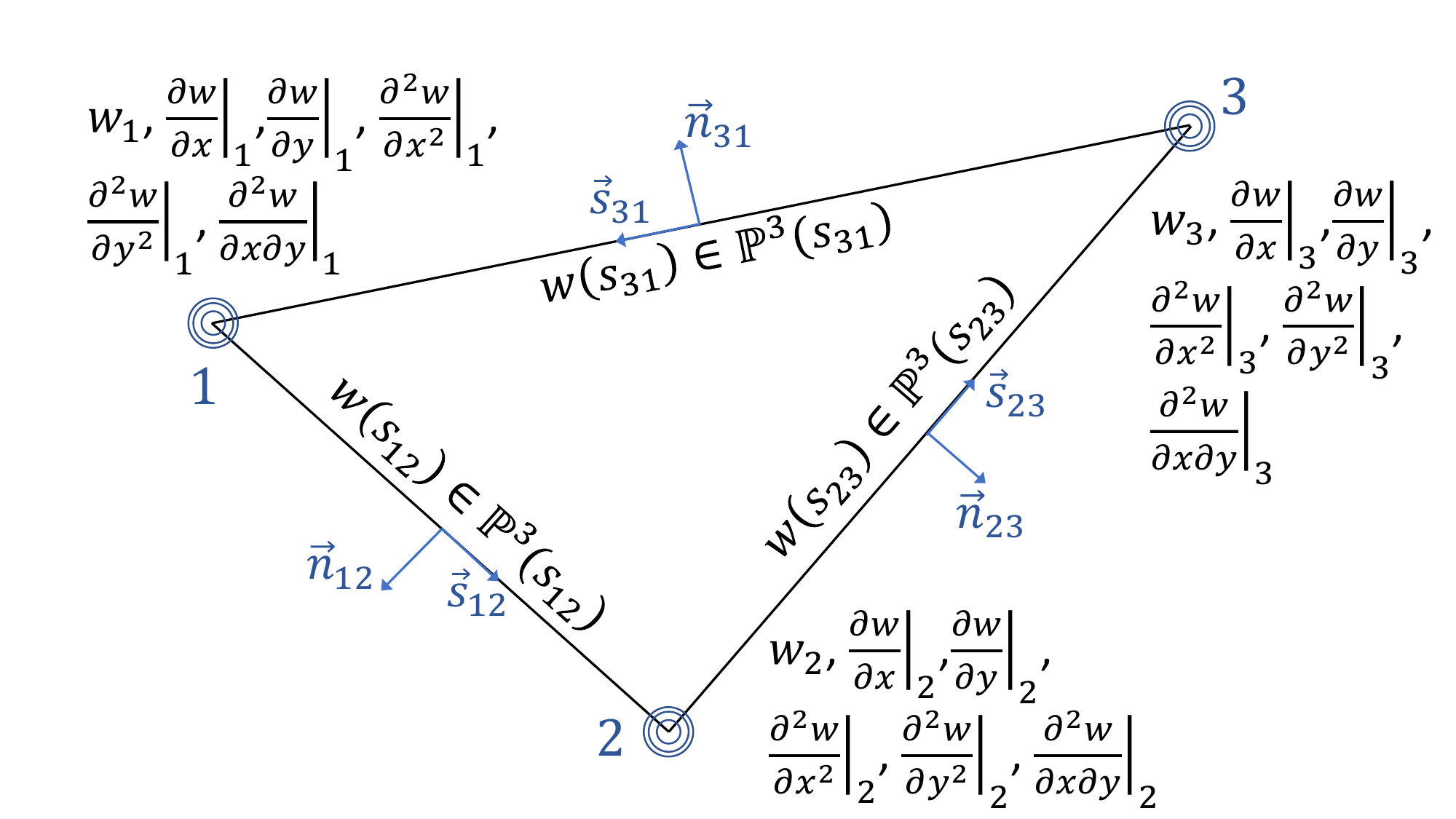}
        \caption{Bell triangle (TUBA3).}
        \label{fig:ps_tuba3}
    \end{minipage}
\end{figure}

 The Argyris triangle or TUBA6, exactly interpolates a 5\textsuperscript{th} order polynomial in area coordinates, which contains all the $L_1^{\alpha}L_2^{\beta}L_3^{\gamma}$ products, such that $\alpha+\beta+\gamma=5$. The out-of-plane displacement is then \cite{Dasgupta1990ARevisited}
\begin{equation}
    \begin{aligned}
        w=&\alpha_1L_1^5+\alpha_2L_2^5+\alpha_3L_3^5+\alpha_4L_1^4L_2+\alpha_5L_1^4L_3\\
        &+\alpha_6L_2^4L_1+\alpha_7L_2^4L_3+\alpha_8L_3^4L_1+\alpha_9L_3^4L_2\\
        &+\alpha_{10}L_1^3L_2^2+\alpha_{11}L_1^3L_3^2+\alpha_{12}L_2^3L_1^2+\alpha_{13}L_2^3L_3^2\\
        &+\alpha_{14}L_3^3L_1^2+\alpha_{15}L_3^3L_2^2+\alpha_{16}L_1^3L_2L_3+\alpha_{17}L_2^3L_1L_3\\
        &+\alpha_{18}L_3^3L_1L_2+\alpha_{19}L_1L_2^2L_3^2+\alpha_{20}L_2L_1^2L_3^2+\alpha_{21}L_3L_1^2L_2^2
    \end{aligned}\label{eq:tuba3_w_monomial}
\end{equation}
As represented in \figref{fig:ps_tuba6}, the Argyris triangle has 6 nodes and 21 DOFs, so all the coefficients in \eqref{eq:tuba3_w_monomial} can be determined uniquely.
\par The Bell triangle is obtained by removing the mid-edge nodes in TUBA6 and by imposing a cubic variation of $\partial{w}/\partial{n}$ along each edge \cite{Bell1969AElement}, such that
\begin{equation}
    \pdv{w}{n_{ij}}\in \mathbb{P}^3(s_{ij}) \mbox{\hspace{7mm} $i,j=\text{1,2,3};\;i\neq j$ }
    \label{eq:ps_tuba3_cubicvar}
\end{equation}
In \Cref{eq:ps_tuba3_cubicvar}, $s_{ij}$ is the $ij^{th}$ edge's coordinate (\Cref{fig:ps_tuba6,fig:ps_tuba3}) and $\mathbb{P}^3(s_{ij})$ is the space of the cubic polynomials along $s_{ij}$. Three of the coefficients in \Cref{eq:tuba3_w_monomial} can thus be expressed in terms of the other 18, by enforcing \Cref{eq:ps_tuba3_cubicvar} for each edge. The remaining coefficients are found by imposing the expression of $w$ or one of its derivatives equal to the nodal quantities
\begin{equation}
    \bm{U_i}^T=
    \begin{bmatrix}
        w\vert_i& \pdv{w}{x}\vert_i& \pdv{w}{y}\vert_i& \pdv[2]{w}{x}\vert_i& \pdv{w}{x}{y}\vert_i& \pdv[2]{w}{y}\vert_i 
    \end{bmatrix}
    \mbox{\hspace{7mm} $i=\text{1,2,3}$ }
    \label{eq:tuba3_dofs}
\end{equation}
\par The components in \Cref{eq:tuba3_dofs} represent the DOFs of the Bell triangle, which can be grouped in the overall DOFs vector as
\begin{equation}
    \bm{U}^T=
    \begin{bmatrix}
        \bm{U_1}^T&\bm{U_2}^T&\bm{U_3}^T
    \end{bmatrix}_{1\times18}
    \label{eq:tuba3_U}
\end{equation}
The out-of-plane displacement can then be written, highlighting the DOFs vector, as
\begin{equation}
    w=
    \begin{bmatrix}
        N_1&N_2&N_3&\dots&N_{18}
    \end{bmatrix}
    \bm{U}=\bm{N} \, \bm{U}
    \label{eq:tuba3_w}
\end{equation}
In \Cref{eq:tuba3_w}, $N_1$ to $N_{18}$ are the TUBA3 shape functions, reported explicitly in Appendix \ref{App:ApA}.

\subsubsection{TUBA3: stiffness matrix and residuals vector}
\par The generalized strain vector is defined for a plate as
\begin{equation}
    \bm{\varepsilon}^T=
    \begin{bmatrix}
        \pdv[2]{w}{x}&\pdv[2]{w}{y}&2\pdv{w}{x}{y}
    \end{bmatrix}
    \label{eq:tuba3_eps}
\end{equation}
After discretization, we can write 
\begin{equation}
    \bm{\varepsilon}=\bm{B} \, \bm{U}
    \label{eq:tuba3_eps_vec}
\end{equation}
where the $\bm{B}$-matrix contains the derivatives of the shape functions with respect to $x$ and $y$. Following the procedure carried out by Dasgupta and Sengupta \cite{Dasgupta1990ARevisited}, $\bm{B}$ can be expressed as the matrix product of other two matrices $\bm{F}$ and $\bm{Q}$, such that
\begin{equation}
    [B]_{3\times18}=\frac{1}{4A^2}[F]_{3\times30}[Q]_{30\times18}
    \label{eq:tuba3_Bmatrix}
\end{equation}
The matrix $\bm{F}$ only contains terms in the area coordinates and reads
\begin{equation}
    [F]_{3x30}=
    \begin{bmatrix}
        [L]^T&[0]_{1\times10}&[0]_{1\times10}\\
        [0]_{1\times10}&[L]^T&[0]_{1\times10}\\
        [0]_{1\times10}&[0]_{1\times10}&[L]^T
    \end{bmatrix}
    \label{eq:tuba3_F}
\end{equation}
with 
\begin{equation}
    [L]^T=
    \begin{bmatrix}
        L_1^3&L_2^3&L_3^3&L_1^2L_2&L_1^2L_3&L_2^2L_1&L_2^2L_3&L_3^2L_2&L_3^2L_2&L_1L_2L_3
    \end{bmatrix}
    \label{eq:tuba3_L}
\end{equation}
$\bm{Q}$ is a matrix which ultimately contains the corners coordinates multiplied together. It can be expressed in terms of three sub-matrices, relatively to the $x$, $y$ and mixed curvatures as
\begin{equation}
    \bm{Q}=
    \begin{bmatrix}
        [Q_{xx}]_{10\times18}\\
        [Q_{yy}]_{10\times18}\\
        [Q_{xy}]_{10\times18}
    \end{bmatrix}
\end{equation}
The detailed components of $\bm{Q}$ are reported as Fortran codes in Dasgupta and Sengupta \cite{Dasgupta1990ARevisited}.
\par The generalized stress vector for a plate is defined as
\begin{equation}
    \bm{\sigma}^T=
    \begin{bmatrix}
        M_{x}&M_{y}&M_{xy}
    \end{bmatrix}
    \label{eq:ps_tuba3_sigma}
\end{equation}
The plate's bending stiffness matrix $\bm{D}$ relates the generalized stresses and strains as
\begin{equation}
\label{eq:plate_D}
    \bm{\sigma}=\bm{D} \, \bm{\varepsilon}
\end{equation} 

\begin{figure}
	\centering
	\includegraphics[width=0.7\textwidth]{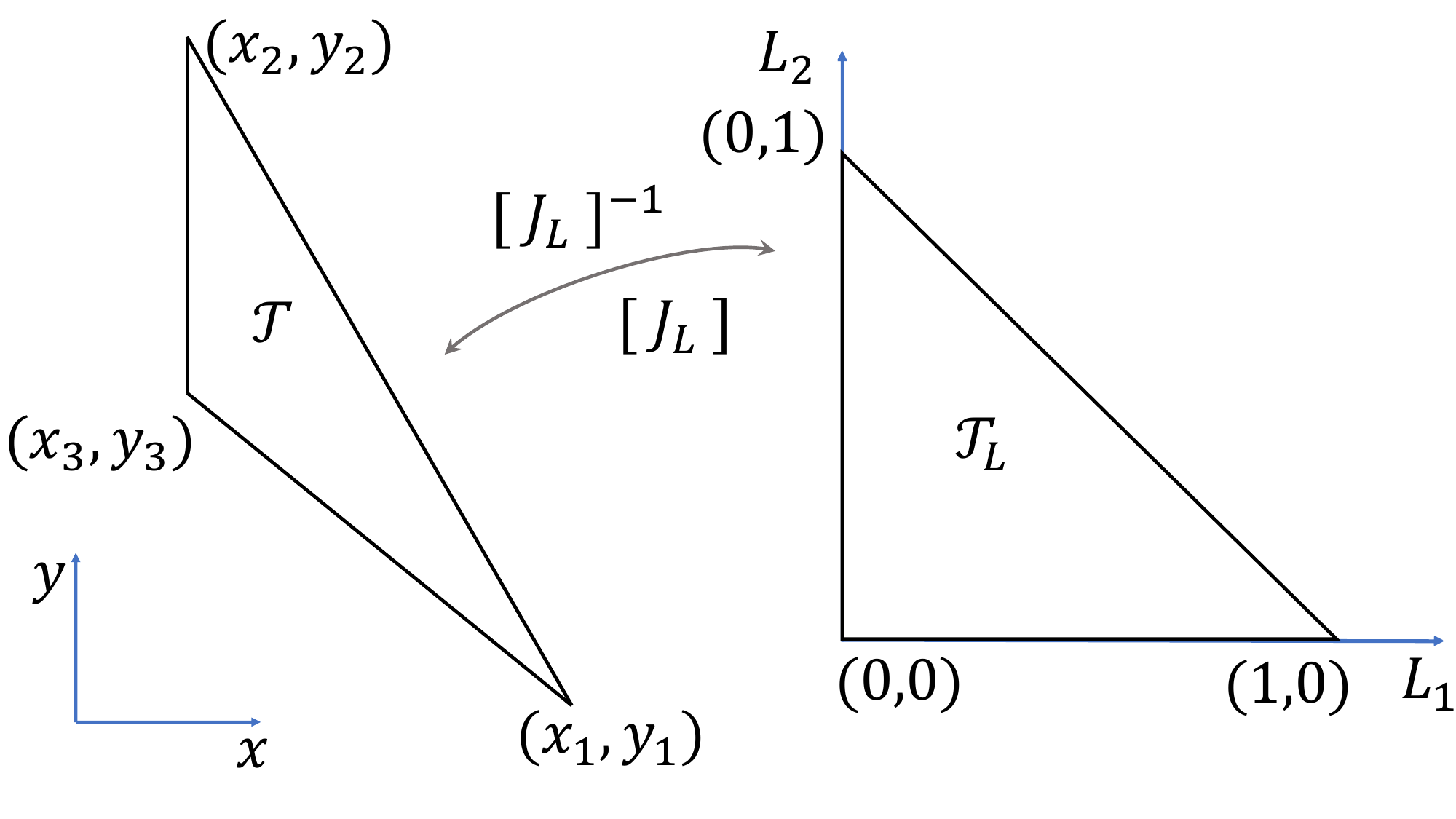}
	\caption{Mapping between a triangle in the physical domain and in the parent triangle.} 
	\label{fig:change_coordinates_x-L}
\end{figure}

\par As indicated in \figref{fig:change_coordinates_x-L}, the configuration of a TUBA element in the physical space is parameterized in the parent domain of the area coordinates. The mapping is expressed by \Cref{eq:x_from_area_coords} and its Jacobian $\bm{J_L}$ is the gradient of the physical coordinates with respect to the parent ones. Integration of the element's stiffness matrix requires the existence of the Jacobian's determinant, which is only possible if $\bm{J_L}$ is a square matrix. For this sake, the area coordinates are reduced from 3 to 2, to match the number of physical coordinates. This is easily achieved, since
\begin{equation}
    L_3=1-L_1-L_2
    \label{eq:ps_reduce_area_coord}
\end{equation}
Substituting Equation \eqref{eq:ps_reduce_area_coord} in the first two equations of \eqref{eq:x_from_area_coords}, allows to have $x$ and $y$ as functions of the sole $L_1$ and $L_2$. The Jacobian is then
\begin{equation}
    \bm{J_L}=
    \begin{bmatrix}
        \pdv{x}{L_1}&\pdv{x}{L_2}\\
        \pdv{y}{L_1}&\pdv{y}{L_2}
    \end{bmatrix}
    =
    \begin{bmatrix}
        x_1-x_3&x_2-x_3\\
        y_1-y_3&y_2-y_3
    \end{bmatrix}
    \label{eq:jacobian}
\end{equation}
and the determinant follows as
\begin{equation}
    \text{det}(\bm{J_L})=(x_1-x_3)(y_2-y_3)-(y_1-y_3)(x_2-x_3) = 2A
    \label{eq:jacobian_det}
\end{equation}
\par \Cref{eq:tuba3_Bmatrix,eq:plate_D} define the TUBA3 stiffness matrix, written as
\begin{equation}
    \bm{K}= \iint_{A}\bm{B}^T \bm{D} \bm{B}\: \text{d}A
    \label{eq:tuba3_K_classical_xy}
\end{equation}
However, $\bm{B}$ contains the shape functions and their derivatives written in area coordinates, therefore the integral over the generic triangle $\mathcal{T}$  is performed instead on the parent triangle $\mathcal{T}_L$. Since
\begin{equation}
    \text{d}A=\text{d}x\,\text{d}y=\text{det}(\bm{J_L})\: \text{d}L_1\, \text{d}L_2
    \label{eq:CE_TUBA3_dA}
\end{equation}
the integral in \Cref{eq:tuba3_K_classical_xy} becomes 
\begin{equation}
    \bm{K}=\int_{0}^{1}\int_{0}^{1-L_1}\bm{B}^T\bm{D}\bm{B}\: \text{det}(\bm{J_L})\: \text{d}L_1\, \text{d}L_2
    \label{eq:tuba3_K_classical_L}
\end{equation}
If \Cref{eq:tuba3_Bmatrix} is substituted in the expression of $\bm{B}$, the integration restricts to the term $\bm{F}^T\bm{F}$, which is a banded matrix, built by repeating the following sub-matrix
\begin{equation}
    \bm{R}=\iint_A \bm{L} \bm{L}^T dA
    \label{eq:tuba3_R}
\end{equation}
three times along the main diagonal. For isotropic materials or balanced composite laminates, $D_{13}, D_{23}, D_{31},$ and $D_{32}$ are zero. This is the case for the DCB problem studied in this work. $\bm{K}$ can then be written as
\begin{multline}
        \bm{K}=\frac{1}{8A^3}\bigg[ D_{11}\bm{Q_{xx}}^T\bm{R}\bm{Q_{xx}}+D_{12}\bm{Q_{xx}}^T\bm{R}\bm{Q_{yy}}+D_{12}\bm{Q_{yy}}^T\bm{R}\bm{Q_{xx}}+\\D_{22}\bm{Q_{yy}}^T\bm{R}\bm{Q_{yy}}+D_{33}\bm{Q_{xy}}^T\bm{R}\bm{Q_{xy}} \bigg]
    \label{eq:tuba3_K_ds}
\end{multline}
\par Numerical integration of $\bm{R}$ can be avoided, by noticing that every integrand in $\bm{R}$ has the form $L_1^aL_2^bL_3^c$. These terms can be integrated in closed form by means of the Eisenberg-Malvern formula \cite{EisenbergOnfinite}, which reads
\begin{equation}
    \iint_A L_1^aL_2^bL_3^c\: \text{d}A = \frac{a!b!c!}{(a+b+c+2)!}2A
    \label{eq:tuba3_eis_mal}
\end{equation}
Analytical integration of the stiffness matrix greatly reduces the CPU time required per-element, making TUBA3 appealing not only due to its regularity but also from an efficiency point of view.
\par The residual vector is expressed as the difference between the external and internal nodal forces:
\begin{equation}
    \bm{f}=\bm{f_{ext}}-\bm{f_{int}}
\end{equation}
A plate element can be subject to distributed surface loads, for which $\bm{f_{ext}}\neq 0$. As an example, Dasgupta and Sengupta derived this vector for a constant applied load \cite{Dasgupta1990ARevisited}. The internal force vector has the usual form:
\begin{equation}
    \bm{f_{int}}=\bm{K}\bm{U}
\end{equation}

\subsection{The TUBA3-compatible cohesive element}
\label{sec:TUBA3-CE}
The TUBA3 plate elements in the previous section will be used to model the composite plies in the DCB problem. In this section, we will develop the CE compatible to the TUBA3 ply elements, named TUBA3-CE. 

\subsubsection{TUBA3-CE: DOFs and kinematics}
\begin{figure}
    \centering
\includegraphics[width=0.8\textwidth]{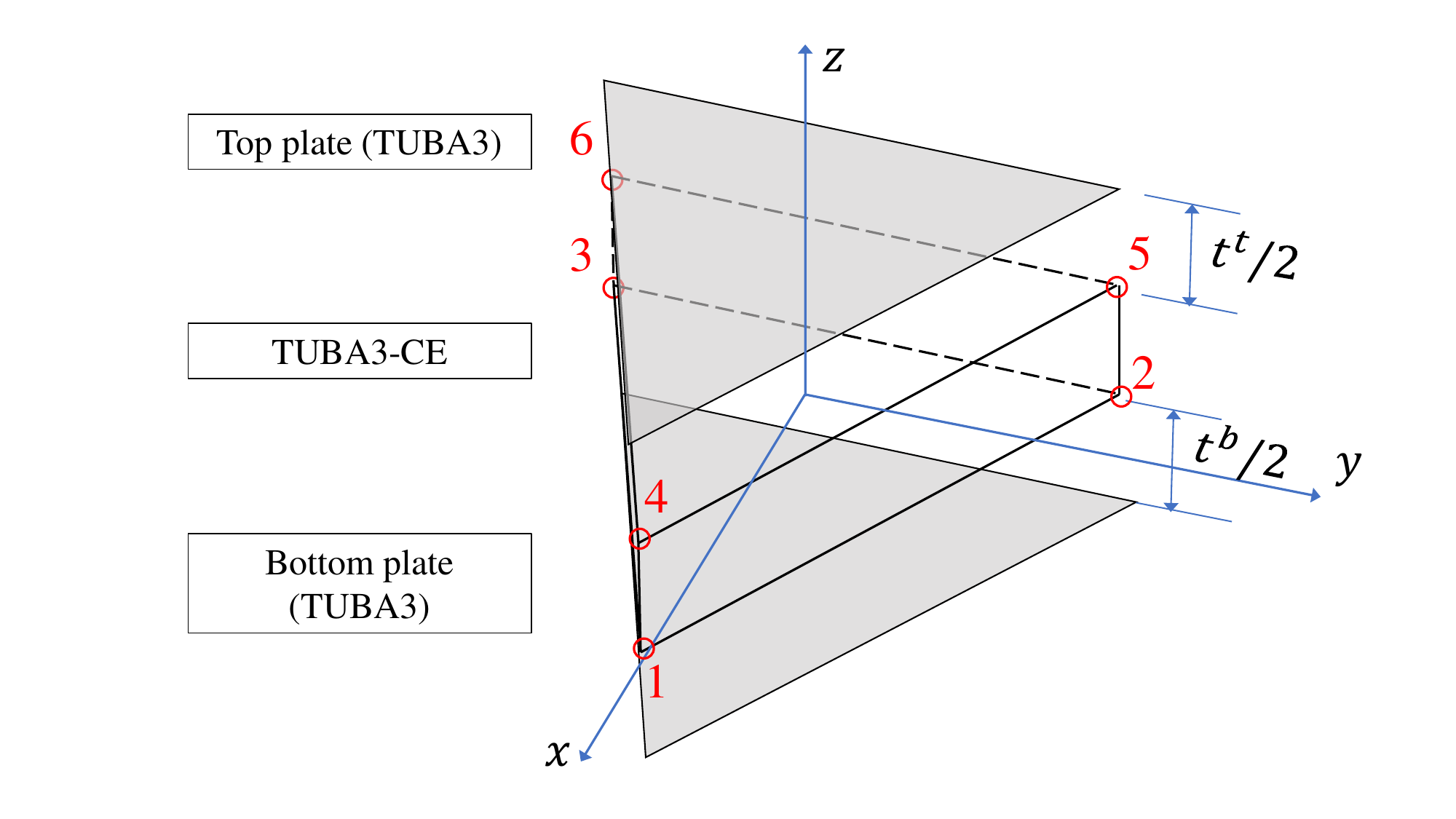}
    \caption{TUBA3 plates and TUBA3-CE: initial configuration.}
    \label{fig:ps_tuba3_assembly}
\end{figure}
The undeformed configuration of TUBA3-CE is illustrated in \figref{fig:ps_tuba3_assembly}. The nodes of the CE share those of the plate elements, which are located at the mid-planes of the plates. The CE's DOFs vector is obtained by stacking those of the bottom and top plate elements. The ordering of the nodes goes from bottom to top, following the right hand rule with respect to the element's normal, which would be along $z$ In \figref{fig:ps_tuba3_assembly}. Referring to \Cref{eq:tuba3_dofs,eq:tuba3_U}, the DOFs vector $\bm{U}$ for TUBA3-CE can be written as
\begin{align}
    \bm{U}^T=&
    \begin{bmatrix}
        U_1&U_2&\dots&U_{18}&U_{19}&\dots&U_{36}
    \end{bmatrix}\nonumber\\
    =&
    \begin{bmatrix}
        w\vert_1& \pdv{w}{x}\vert_1&\dots&\pdv[2]{w}{y}\vert_3& w\vert_4& \pdv{w}{x}\vert_4&\dots&\pdv[2]{w}{y}\vert_6
    \end{bmatrix}
    \label{eq:tuba3CE_U}
\end{align}
\begin{figure}[p]
    \centering
    \includegraphics[width=0.9\textwidth]{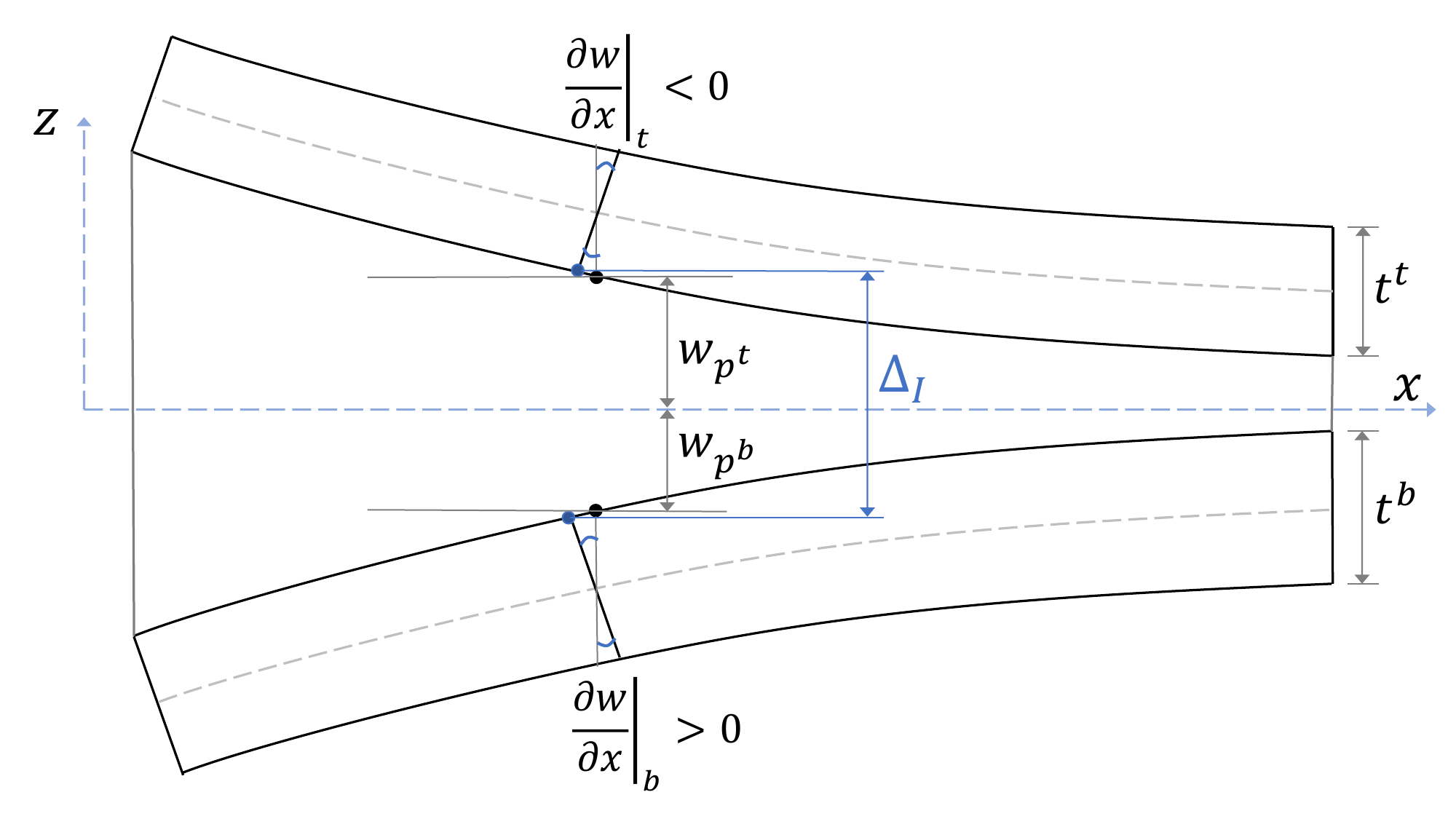}
    \caption{Mode I opening. The undeformed CE initially coincides with the $x$-axis.}
    \label{fig:ps_tuba3CE_mode1}
\end{figure}
\begin{figure}[p]
    \centering
    \includegraphics[width=0.9\textwidth]{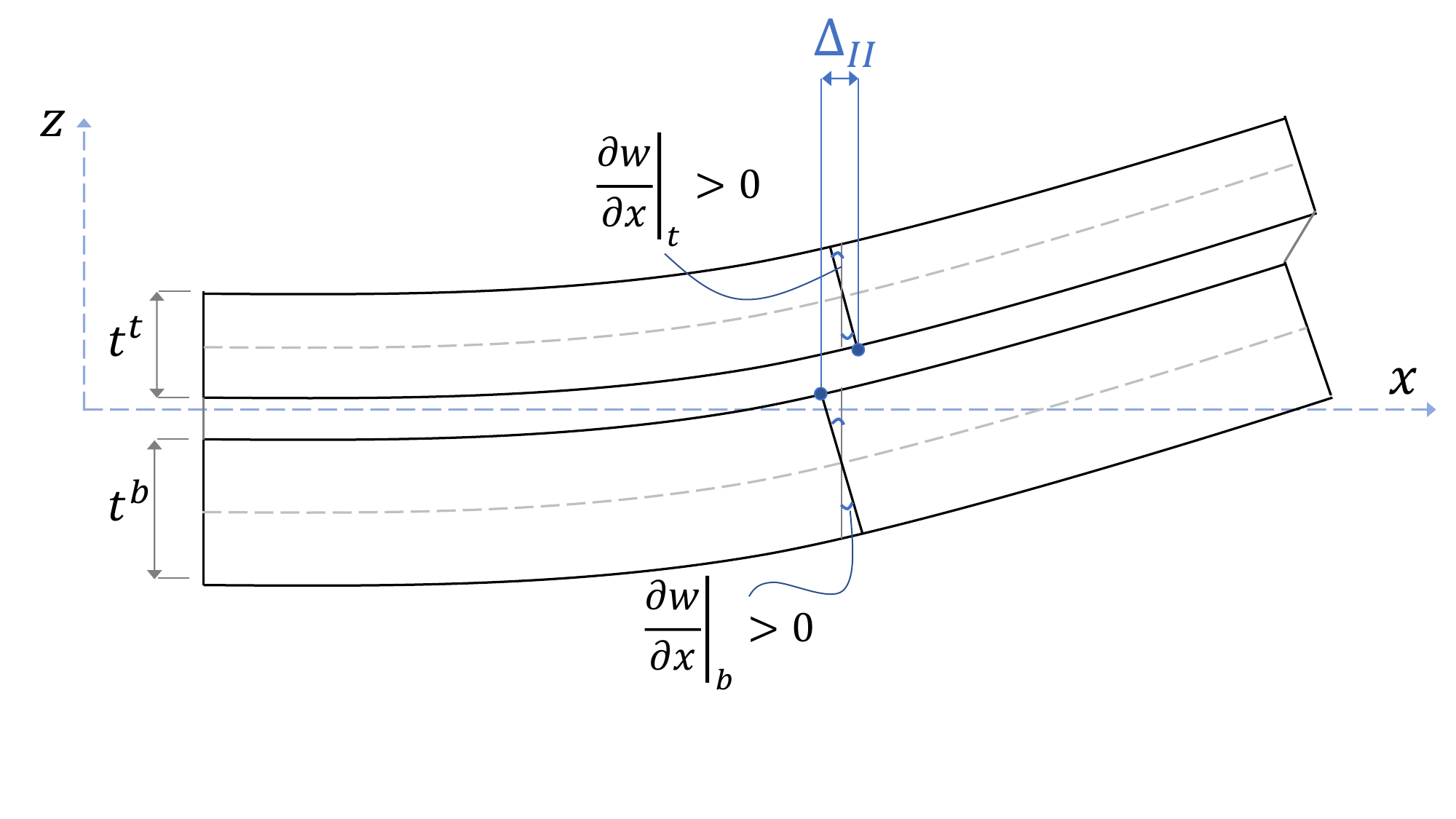}
    \caption{Mode II opening. The undeformed CE initially coincides with the $x$-axis.}
    \label{fig:ps_tuba3CE_mode2}
\end{figure}
\par From here onwards, the indices $CE^b$ and $CE^t$ refer to quantities relative to the bottom and top CE faces, respectively. Similarly, $p^b$ and $p^t$ refer to those relative to the mid-planes of bottom and top plate elements, respectively. The terms $t^b$ and $t^t$ indicate the thicknesses of the bottom and top plates, respectively. 
\figref{fig:ps_tuba3CE_mode1} shows how rotations contribute slightly to the Mode I opening of the faces of the CE. For geometrical linear analysis, such contributions are negligible. The mode I opening is then expressed only by the deflections of the mid-planes of the plate elements, hence
\begin{equation}
    \Delta_I=-w_{p^b}+w_{p^t}
    \label{eq:tuba3CE_delta_1}
\end{equation}

\par \figref{fig:ps_tuba3CE_mode2} shows the deformation of TUBA3-CE in pure mode II. Since the TUBA3 plate element only deforms in bending and does not have stretching DOFs, opening in mode II is caused only by the rotations of bottom and top plates. Therefore
\begin{equation}
   \Delta_{II}=-u_{CE^b}+u_{CE^t}
\end{equation}
where
\begin{eqnarray}
   u_{CE^b}=&-\frac{t^b}{2}\sin(\pdv{w_{p^b}}{x})\\
   u_{CE^t}=&\frac{t^t}{2}\sin(\pdv{w_{p^t}}{x})
\end{eqnarray}
Assuming small rotations, the sines can be approximated at the first order as
\begin{eqnarray}
   &\sin(\pdv{w_{p^b}}{x}) \approx \pdv{w_{p^b}}{x} \nonumber\\
   &\sin(\pdv{w_{p^t}}{x}) \approx \pdv{w_{p^t}}{x} \nonumber
\end{eqnarray}
It follows that $\Delta_{II}$ is expressed as
\begin{equation}
    \Delta_{II}= \frac{t^b}{2}\pdv{w_{p^b}}{x} + \frac{t^t}{2}\pdv{w_{p^t}}{x}
    \label{eq:tuba3CE_delta_2}
\end{equation}
\par The mode III opening is found from analogous kinematics in the $yz$-plane. Hence
\begin{equation}
    \Delta_{III}= \frac{t^b}{2}\pdv{w_{p^b}}{y} + \frac{t^t}{2}\pdv{w_{p^t}}{y}
    \label{eq:tuba3CE_delta_3}
\end{equation}
\par \Cref{eq:tuba3CE_delta_1,eq:tuba3CE_delta_2,eq:tuba3CE_delta_3} can be expressed in terms of the element DOFs. The vector of the DOFs can be written in the form
\begin{equation}
    \bm{U}^T=
    \begin{bmatrix}
        \bm{U}^{\bm{b}\,T}&\bm{U}^{\bm{t}\,T}
    \end{bmatrix}_{1\times36}
    \label{eq:tuba3CE_Ub_Ut}
\end{equation}
which highlights the DOFs belonging respectively to upper and lower face. Recalling \Cref{eq:tuba3_w,eq:tuba3CE_delta_1}, the mode I opening becomes
\begin{equation}
    \Delta_I= -\bm{N} \bm{U^b}+\bm{N} \bm{U^t}=
    \begin{bmatrix}
        -[N]_{1\times18}&[N]_{1\times18}
    \end{bmatrix}
    \bm{U}=\bm{B_I} \, \bm{U}
    \label{eq:tuba3CE_delta_1_U}
\end{equation}
\par The opening in mode II requires the $x$-derivatives of the shape functions, since
\begin{equation}
   \pdv{w}{x}=\left[\pdv{\bm{N}}{x}\right] \bm{U}
\end{equation}
The shape functions are expressed in the area coordinates $L_1$,$L_2$,$L_3$, thus the $x$ and $y$ derivatives are computed using the chain rule of differentiation. Recalling \Cref{eq:area_coords_from_x},
\begin{equation}
    \left[\pdv{\bm{N}}{x}\right]_{1\times18}=\frac{1}{2A}
    \begin{bmatrix}
        b_1&b_2&b_3
    \end{bmatrix}
    \begin{bmatrix}
        \left[\pdv{\bm{N}}{L_1}\right]_{1\times18}\\
        \left[\pdv{\bm{N}}{L_2}\right]_{1\times18}\\
        \left[\pdv{\bm{N}}{L_3}\right]_{1\times18}
    \end{bmatrix}
    \label{eq:tuba3CE_dndx}
\end{equation}
Once the derivatives in \Cref{eq:tuba3CE_dndx} are obtained, it is possible to express the mode II opening in terms of the DOFs, as
\begin{equation}
    \Delta_{II}=
    \begin{bmatrix}
        \frac{t^b}{2}\left[\pdv{\bm{N}}{x}\right] & \frac{t^t}{2}\left[\pdv{\bm{N}}{x}\right]
    \end{bmatrix}
    \bm{U}=\bm{B_{II}} \, \bm{U}
    \label{eq:tuba3CE_delta_2_U}
\end{equation}
\par Starting from \cref{eq:tuba3CE_delta_3} and deriving with respect to $y$, the opening in mode III can be written as
\begin{equation}
    \Delta_{III}=
    \begin{bmatrix}
        \frac{t^b}{2}\left[\pdv{\bm{N}}{y}\right] & \frac{t^t}{2}\left[\pdv{\bm{N}}{y}\right]
    \end{bmatrix}
    \bm{U}=\bm{B_{III}} \, \bm{U}
    \label{eq:tuba3CE_delta_3_U}
\end{equation}
where, this time
\begin{equation}
    \left[\pdv{\bm{N}}{y}\right]_{1\times18}=\frac{1}{2A}
    \begin{bmatrix}
        c_1&c_2&c_3
    \end{bmatrix}
    \begin{bmatrix}
        \left[\pdv{\bm{N}}{L_1}\right]_{1\times18}\\
        \left[\pdv{\bm{N}}{L_2}\right]_{1\times18}\\
        \left[\pdv{\bm{N}}{L_3}\right]_{1\times18}
    \end{bmatrix}
    \label{eq:tuba3CE_dndy}
\end{equation}
\par \Cref{eq:tuba3CE_delta_1_U,eq:tuba3CE_delta_2_U,eq:tuba3CE_delta_3_U} can be finally assembled together to form the $\bm{B}$-matrix of TUBA3-CE
\begin{equation}
    \begin{bmatrix}
        \Delta_I\\
        \Delta_{II}\\
        \Delta_{III}
    \end{bmatrix}
    =\begin{bmatrix}
        \bm{B_I}\\
        \bm{B_{II}}\\
        \bm{B_{III}}
    \end{bmatrix}_{3\times36}
    \bm{U}=\bm{B}\,\bm{U}
    \label{eq:tuba3CE_B}
\end{equation}

\subsubsection{TUBA3-CE: constitutive relation}
The constitutive law of a CE links tractions $\bm{\tau}$ and separations $\bm{\Delta}$. The constitutive law of TUBA3-CE is the same as that of the standard CE. This is defined by the constitutive matrix $\bm{D}$:
\begin{equation}
    \bm{\tau}=\bm{D}\,\bm{\Delta}
\end{equation}
The $\bm{D}$-matrix usually assumes the following form:
\begin{equation}
    \bm{D}=
    \begin{bmatrix}
        (1-d_I)K&0&0\\
        0&(1-d)K&0\\
        0&0&(1-d)K
    \end{bmatrix}
    \label{eq:D_CE_std}
\end{equation}
where $K$ is the penalty stiffness, $d_I$ and $d$ are the damage variables for opening (Mode I) and shear (Mode II and III) delamination, respectively. The damage variable in mode I is distinguished from that in shear, in order to avoid interpenetration of opposite crack surfaces
\begin{equation}
    \begin{cases}
        d_I=d  &\mbox{\hspace{7mm} if $\Delta_I>0$ }\\
        d_I=0  &\mbox{\hspace{7mm} if $\Delta_I\leq0$ }
    \end{cases}
    \label{eq:d_prime}
\end{equation}
The bilinear cohesive law is used, with the penalty stiffness set as \cite{Turon2007AnModels}:
\begin{equation}
    K=50\,\frac{E_3}{t}
    \label{eq:tuba3CE_modeI_K}
\end{equation}
where $E_3$ is the out-of-plane Young's modulus and $t$ is the thickness of the laminate. Without loss of generality, the cohesive law in this work pertains to Mode I only, as this study focuses on tackling this case which comes with the most stringent requirement on the mesh density in the cohesive zone.

\subsubsection{TUBA3-CE: stiffness matrix and residuals vector}
\label{ssec:TUBA3-CE_K}
The element stiffness matrix has the usual form seen in \Cref{eq:tuba3_K_classical_xy}. The parent domain of TUBA3-CE is the same flat triangle used for TUBA3, thus the two elements share the same coordinates mapping and Jacobian. Referring to the $\bm{B}$ and $\bm{D}$ matrices just derived, the matrix $\bm{K}$ for TUBA3-CE reads
\begin{equation}
    \bm{K}= \int_{0}^{1}\int_{0}^{1-L_1}\bm{B}^T\bm{D}\bm{B}\: \text{det}(\bm{J_L})\: \text{d}L_1 \,\text{d}L_2
    \label{eq:tuba3-CE_K}
\end{equation}
where
\begin{equation*}
    \text{det}(\bm{J_L})=2A
\end{equation*}
As no external distributed loads are applied on the surfaces of the CE, the residual vector at every iteration is simply
\begin{equation}
    \bm{f}=-\bm{K}\bm{U}
\end{equation}
\par For the TUBA3 plate element, the integral in \Cref{eq:tuba3-CE_K} could be solved analytically, using the Eisenberg-Malvern formula. The same does not always hold for TUBA3-CE. Since the damage variable can change throughout the element's domain, the cohesive element integration does not necessarily reduce to terms such as those in \Cref{eq:tuba3_eis_mal}. Although $\bm{K}$ could be integrated analytically when the damage is homogeneous, this would require to rewrite $\bm{B}$ in a form similar to \Cref{eq:tuba3_Bmatrix} and to isolate the Eisenberg-Malvern terms. This operation is error-prone and the final formulation hard to verify, therefore numerical integration is adopted for all TUBA3-CEs, regardless of their damage state. Gaussian quadrature is chosen over the Newton-Cotes scheme, as the former achieves equal degrees of accuracy with fewer integration points.

\subsubsection{Numerical integration over triangular domains}
\par The integral of a generic function $g(L_1,L_2)$, defined over the parent triangle $\mathcal{T}_L$, is approximated as \cite{deng2010quadrature,cowper1973gaussian}
\begin{equation}
    \iint_{\mathcal{T}_L}g(L_1,\,L_2)\:\text{d}L_1\,\text{d}L_2 \approx \frac{1}{2}\sum_{i=1}^{N_{IP}} w_i\, g(L_{1,i},\,L_{2,i})
    \label{eq:tuba3-CE_int_tri}
\end{equation}
where $N_{IP}$ is the overall number of integration points and $w_i$, $L_{1,i}$, $L_{2,i}$ are respectively the weight and area coordinates of the $i$\textsuperscript{th} integration point. In the literature, several works focused on finding precise estimates of $w_i$, $L_{1,i}$, $L_{2,i}$ for Gaussian quadrature of degrees up to 20 \cite{hammer1956numerical,irons1966engineering,felippa1968refined,cowper1973gaussian,dunavant1985high}. The values used in this research are taken from the paper of Cowper \cite{cowper1973gaussian}.

\subsubsection{Sub-domain integration}
\begin{figure}
	\centering
	\includegraphics[width=0.8\textwidth]{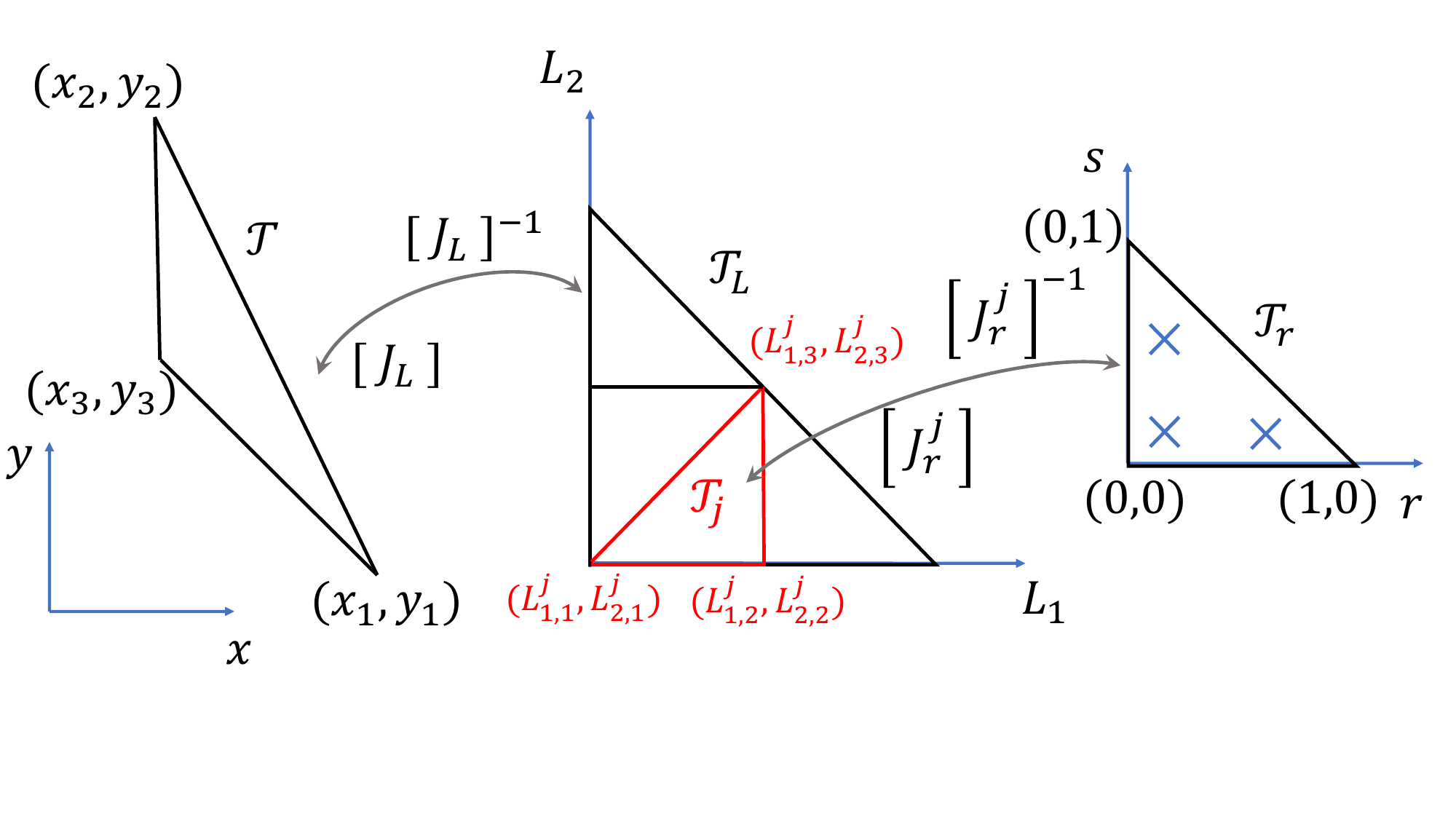}
	\caption{Different mappings between physical and parent domains. A function over each sub-triangle in the $L$-domain can be integrated through Gaussian quadrature in the $r$-domain. Generic IPs in the $r$-domain are highlighted.}  
	\label{fig:change_coordinates_x-L-r}
\end{figure}
Large cohesive elements show highly non-linear stress and damage distributions when crossed by the cohesive zone. If these elements are given an insufficient number of integration points, the FE solution can be sensitive to instabilities and the Newton-Raphson procedure may diverge.
\par An easy way to increase the density of IPs, that does not require the knowledge of formulas for high degrees of quadrature accuracy, is to use a sub-domain integration scheme. The idea is to split the integral over the parent domain of coordinates $L_1$, $L_2$, $L_3$ (from now on called the \textit{$L$-domain}) in multiple integrals over $N_{sd}$ sub-triangles. Each of these sub-triangles or sub-domains is then integrated with Gaussian quadrature over a third domain of coordinates $r$,$s$,$t$, named the \textit{$r$-domain}. \figref{fig:change_coordinates_x-L-r} shows the three different domains (physical, $L$-domain, $r$-domain), the partition in sub-triangles and the location of three Gaussian IPs in the $r$-domain.
\par The additive property allows to write the integral over $\mathcal{T}_L$ as the sum of the integrals over a set of sub-triangles $\mathcal{T}_j$ as follows
\begin{equation}
    \iint_{\mathcal{T}_L}g(L_1,\,L_2)\:\text{d}L_1\,\text{d}L_2=\sum_{j=1}^{N_{sd}}\iint_{\mathcal{T}_j}g(L_1^j,\,L_2^j)\:\text{d}L_1\,\text{d}L_2
    \label{eq:tuba3-CE_integral_additive}
\end{equation}
\par The mapping between $\mathcal{T}_j$ and $\mathcal{T}_L$ is defined as
\begin{equation}
    \begin{cases}
        L_1^j&=r\,L_{1,1}^j + s\,L_{1,2}^j + t\,L_{1,3}^j\\
        L_2^j&=r\,L_{2,1}^j + s\,L_{2,2}^j + t\,L_{2,3}^j\\
        1&=r+s+t
    \end{cases}
    \label{eq:tuba3-CE_L_from_r}
\end{equation}
In \Cref{eq:tuba3-CE_L_from_r}, the notation $L_{m,n}^j$ indicates the $m$\textsuperscript{th} area coordinate of the $n$\textsuperscript{th} vertex of the $j$\textsuperscript{th} sub-triangle in the L-domain. By comparison of \Cref{eq:tuba3-CE_L_from_r,eq:x_from_area_coords} it is evident how the above mapping is again a linear transformation in area coordinates, just like the one between the physical domain and the $L$-domain. It follows that the Jacobian for the $L-r$ mapping is
\begin{equation}
    \bm{J_r^j}=
    \begin{bmatrix}
        \pdv{L_1^j}{r}&\pdv{L_1^j}{s}\\
        \pdv{L_2^j}{r}&\pdv{L_2^j}{s}
    \end{bmatrix}
    =
    \begin{bmatrix}
        L_{1,1}^j-L_{1,3}^j&L_{1,2}^j-L_{1,3}^j\\
        L_{2,1}^j-L_{2,3}^j&L_{2,2}^j-L_{2,3}^j
    \end{bmatrix}
    \label{eq:tuba3-CE_jacobian_r}
\end{equation}
The determinant of $\bm{J_r^j}$ is then
\begin{equation}
    \text{det}(\bm{J_r^j})=(L_{1,1}^j-L_{1,3}^j)(L_{2,2}^j-L_{2,3}^j)-(L_{2,1}^j-L_{2,3}^j)(L_{1,2}^j-L_{1,3}^j) = \text{2}A_{\mathcal{T}_j}
    \label{eq:tuba3-CE_jacobian_r_det}
\end{equation}
\par The integral over the triangle $\mathcal{T}_j$ can now be performed in the $r$-domain as
\begin{equation}
    \iint_{\mathcal{T}_j}g(L_1^j,\,L_2^j)\:\text{d}L_1\,\text{d}L_2=\iint_{\mathcal{T}_r}g(L_1^j(r,s),\,L_2^j(r,s))\,\text{det}(\bm{J_r^j})\:\text{d}s\,\text{d}r
    \label{eq:tuba3-CE_integral_in_r}
\end{equation}
\Cref{eq:tuba3-CE_integral_in_r} can be integrated numerically via Gaussian quadrature with $N_{IP}$ integration points, hence
\begin{equation}
    \iint_{\mathcal{T}_r}g(L_1^j(r,s),\,L_2^j(r,s))\,\text{det}\bm{J_r^j}\:\text{d}s\,\text{d}r = \frac{1}{2}\sum_{i=1}^{N_{IP}} w_i\, g(L_1^j(r_i,s_i),\,L_2^j(r_i,s_i))\,\text{det}(\bm{J_r^j})
    \label{eq:tuba3-CE_integral_in_r_num}
\end{equation}
\par The integral over the entire parent triangle $\mathcal{T}_L$ is given by the contribution of all the sub-triangles $\mathcal{T}_j$ and reads
\begin{equation}
    \iint_{\mathcal{T}_L}g(L_1,\,L_2)\:\text{d}L_1\,\text{d}L_2=\frac{1}{2}\sum_{j=1}^{N_{sd}}\sum_{i=1}^{N_{IP}} w_i\, g(L_1^j(r_i,s_i),\,L_2^j(r_i,s_i))\,\text{det}(\bm{J_r^j})
    \label{eq:tuba3-CE_integral_in_L_num}
\end{equation}

\section{Validation}
\label{sec:VL}
In order to assess the accuracy and performance of the proposed method, the assembly of TUBA3 and TUBA3-CE was tested on the mode I DCB problem \cite{KruegerBenchmark}. The benchmark solution provided in this work and the analytical curve obtained with the Corrected Beam Theory were used as references for comparison.

\subsection{Benchmark problem description}
\begin{figure}
	\centering
	\includegraphics[width=0.9\textwidth]{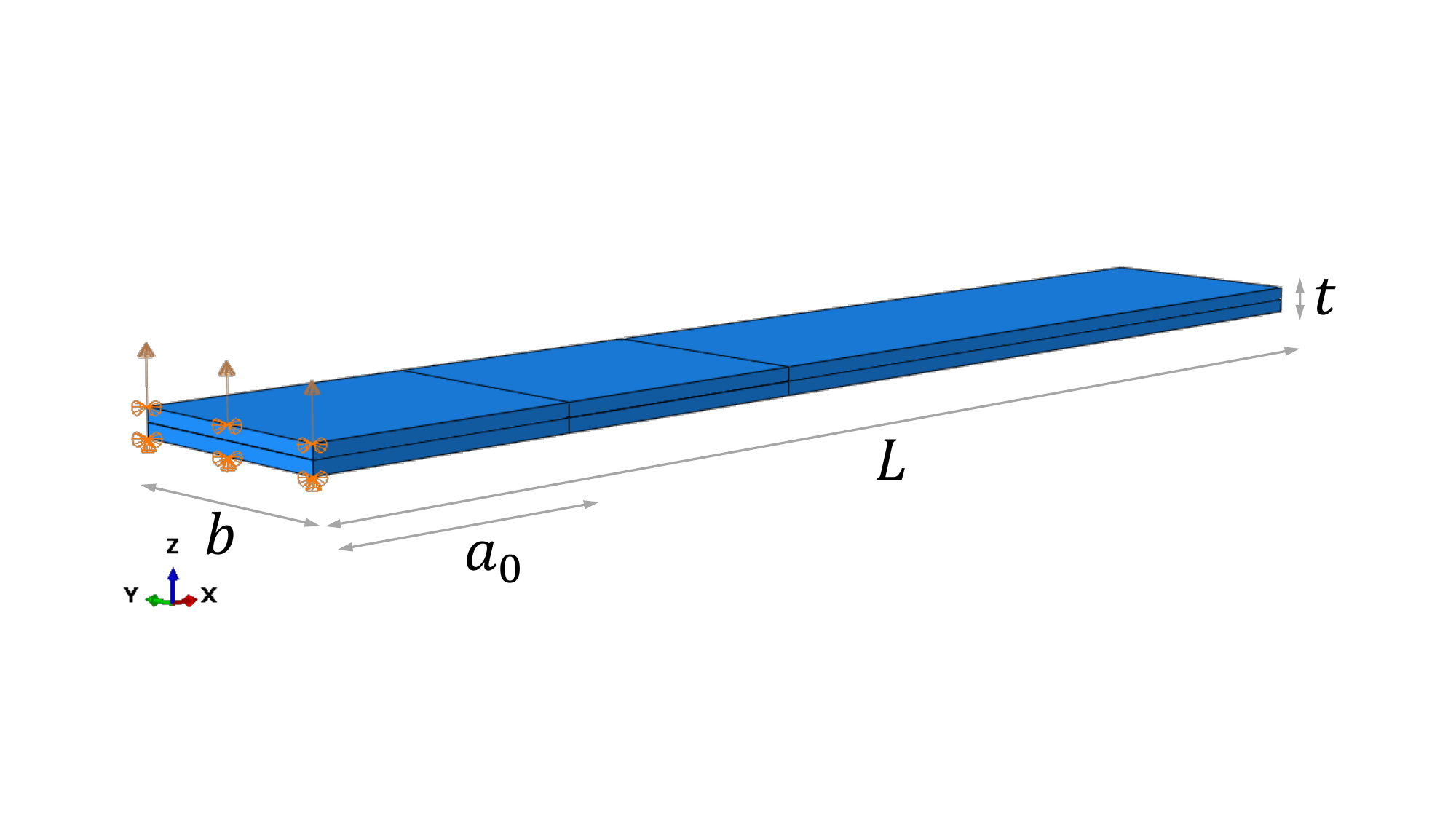}
	\caption{DCB specimen: geometry, loading and boundary conditions. $L=150$~mm, $b=25$~mm, $t=3$~mm, $a_0=30.5$~mm.} 
	\label{fig:vl_geom}
\end{figure}
The DCB specimen is portrayed in \figref{fig:vl_geom} in its undeformed configuration. The dimensions of the specimen and the initial crack size are shown, as well as the boundary conditions and applied loading velocity of the numerical model. The laminate is unidirectional with 24 plies of T300/1076 graphite-epoxy prepreg. Table \ref{tab:vl_DCB_data} summarizes the values of the material properties.
\begin{table}
\centering
\begin{tabular}{cccc}
\hline
\textbf{Material} &   & T300/1076   &      \\
\hline
                                     & $E_{xx}$ & 139.4 & GPa \\
Young's moduli                       & $E_{yy}$ & 10.16 & GPa \\
                                     & $E_{zz}$ & 10.16 & GPa \\
\hline
                                     & $\nu_{xy}$ & 0.3 &    \\
Poisson's ratios                     & $\nu_{xz}$ & 0.3 &   \\
                                     & $\nu_{yz}$ & 0.436 &   \\
\hline
                                     & $G_{xy}$ & 4.6 & GPa      \\
Shear moduli                         & $G_{xz}$ & 4.6 & GPa      \\
                                     & $G_{yz}$ & 3.54 & GPa     \\
\hline
Fracture                             & $G_{I,c}$& 0.170 & kJ/m\textsuperscript{2}  \\
toughnesses                          & $G_{II,c}$& 0.494 & kJ/m\textsuperscript{2}  \\
\hline
Material                             & $\tau_I^0$& 30    & MPa    \\
strengths                            & $\tau_{II}^0$& 50    & MPa    \\
\hline
B-K coefficient                      & $\eta$       & 1.62 &         \\
\hline
\end{tabular}
\caption{DCB material data.}
\label{tab:vl_DCB_data}
\end{table}


\subsection{Description of the reference numerical model}
The following gives an overview of the main modeling choices and features adopted in building the Abaqus reference model. 
\par The entire preprocessing phase was carried out in Abaqus CAE. The model consists of two parts, corresponding to the two sublaminates. Only the bottom part/sublaminate was modelled directly, while the top one was created as an exact copy. The two parts were then stacked together at the assembly level. A partition as shown in \figref{fig:vl_geom} was done to isolate the crack propagation region, which required a mesh finer than for the rest of the specimen. The substrates were assigned a composite layup property with the material data provided in Table \ref{tab:vl_DCB_data}. Since the laminate is unidirectional, a single ply with half the laminate's thickness could be assigned to be the layup of each part. The element type was chosen as the linearly interpolated brick element with incompatible modes or C3D8I, in Abaqus nomenclature.
\par The cohesive interface was modeled by defining cohesive contact (CC) between the two adjacent substrate surfaces, that extend from the end of the precrack to the end of the specimen. If the interface is negligibly thin, CC can be used in place of CEs and it usually offers improved computational performances \cite{Zhang2015SimulatingContact}. The quadratic criterion \cite{Cui1992QuadrCrit} was used for damage initiation and the B-K relation as crack propagation condition. In order to avoid numerical instabilities during the fracture process, a viscosity coefficient $\eta_v$ was used and set equal to 10\textsuperscript{-5} s. This value was found sufficient for a stable solution and small enough to not pollute it with spurious damping.
\par All the analyses were run with the full Newton Raphson method. The boundary conditions for the reference analyses constrained the translations in all directions for the bottom left edge of the specimen and in the $x$ and $y$ directions for the top left one, as presented in \figref{fig:vl_geom}. Loading was imposed on the top left edge for a total displacement of 4 mm along z.

\subsection{Mesh convergence study of the reference model}
\begin{figure}
	\centering
	\includegraphics[width=0.9\textwidth]{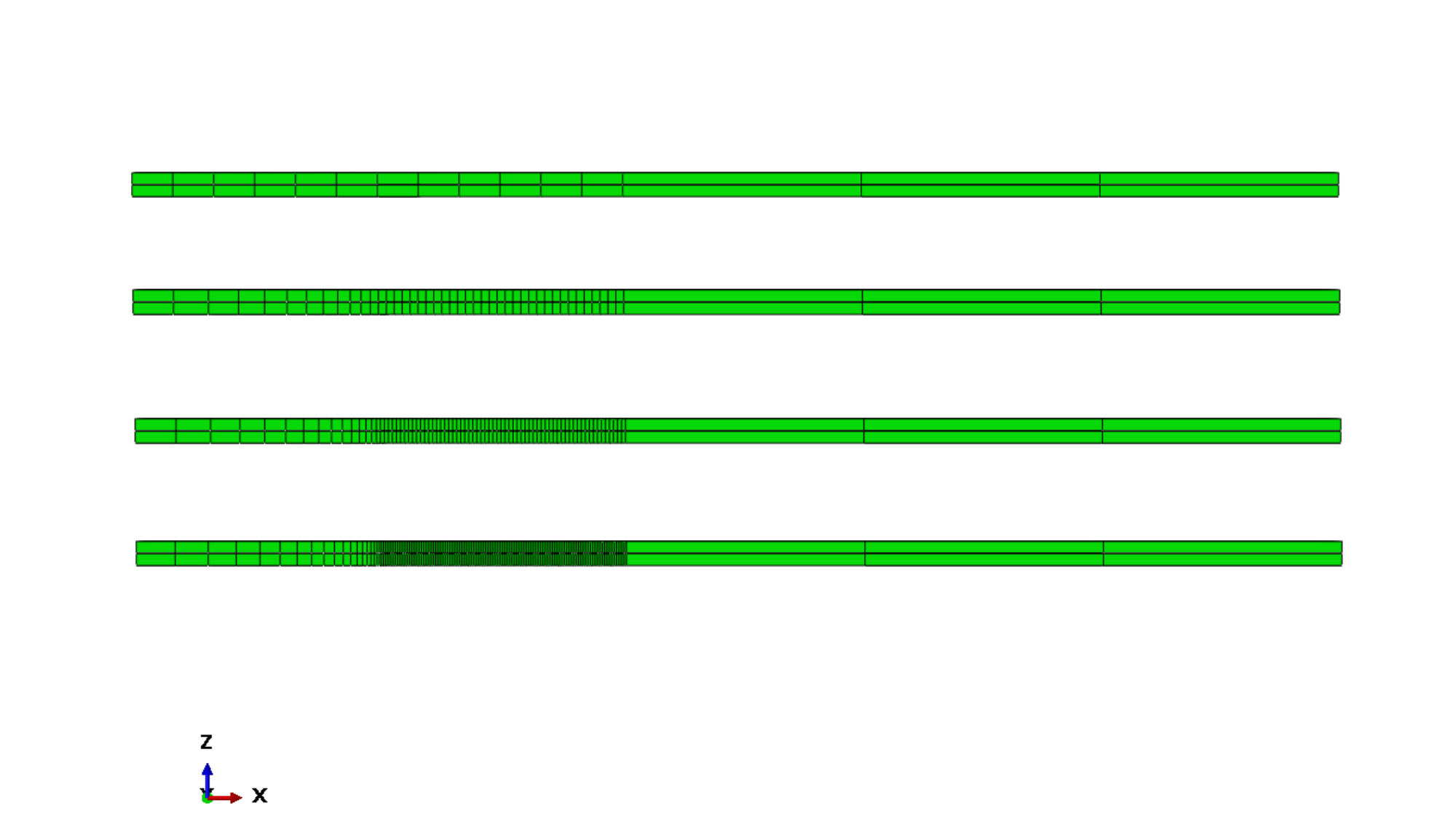}
	\caption{Four meshes of different element sizes in the crack propagation region. The lengths are, from top to bottom: 5mm, 1 mm, 0.5 mm and 0.25 mm.} 
	\label{fig:vl_ref_meshes}
\end{figure}
Using the model just detailed, four different mesh sizes were studied. These are represented in \figref{fig:vl_ref_meshes} for a section view in the $xz$-plane. In all three cases, 1 element was used along the thickness direction and 25 elements covered the width. The meshes differed only for the elements size used in the fracture region, where the element lengths are respectively of 5, 1, 0.5 and 0.25 mm.
\begin{figure}
	\centering
	\includegraphics[width=\textwidth]{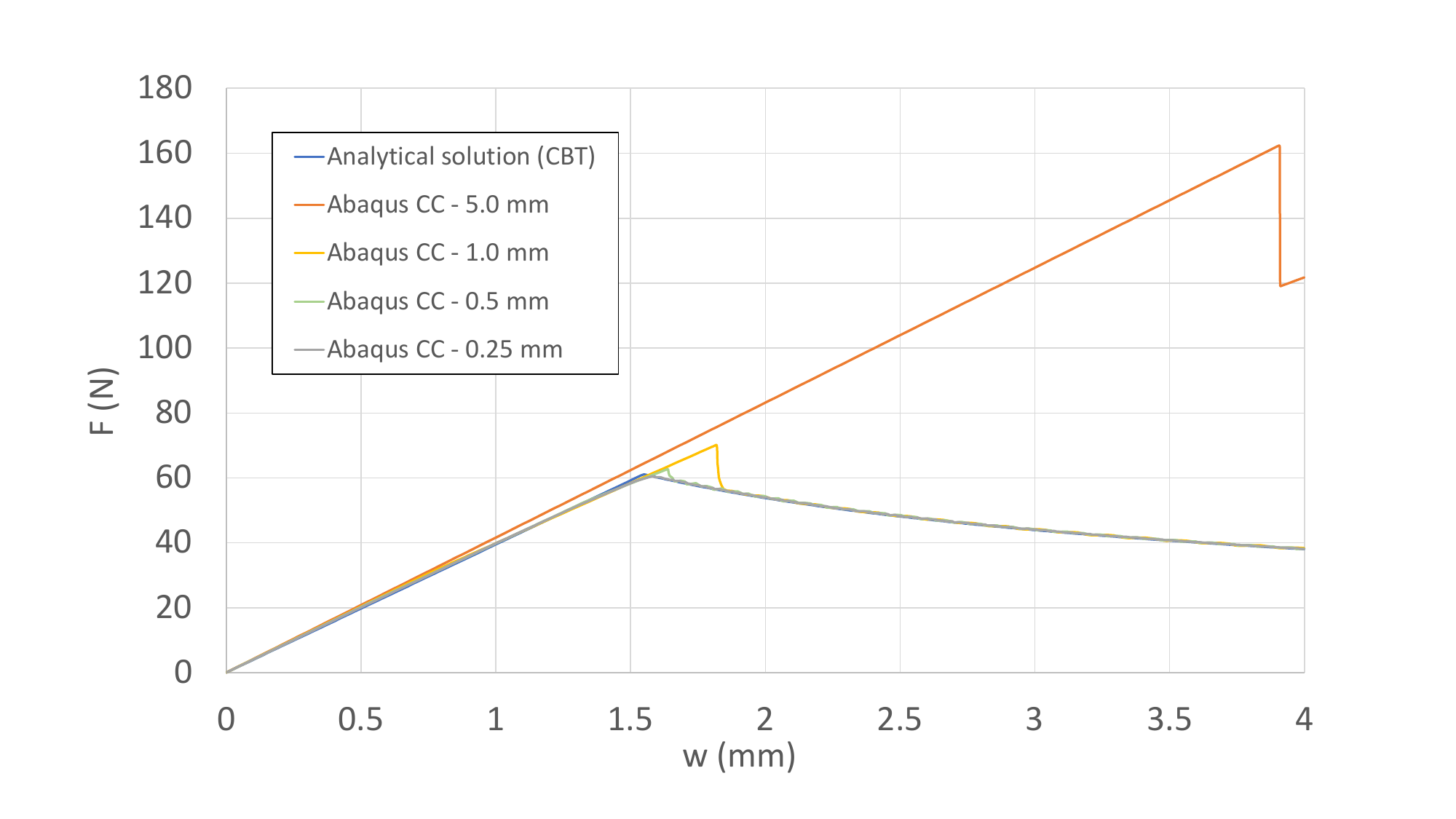}
	\caption{DCB test results obtained using the Abaqus reference model for four different mesh sizes. The Corrected Beam Theory solution is also plotted.} 
	\label{fig:vl_ref_results}
\end{figure}
\par \figref{fig:vl_ref_results} shows the load-displacement curves for the four meshes discussed. These are plotted together with the analytical solution derived from the Corrected Beam Theory. The coarsest mesh of 5 mm misses the limit point completely, as no material damage occurs before reaching a load of 160 N. When the element size is reduced to 1 mm, the peak force is still overpredicted by 15\% the CBT value. The overshoot reduces to 2.6\% when the 0.5 mm case is considered. The finest mesh of 0.25 mm practically overlaps the analytical curve. The critical load and displacement for this element size are respectively 60.5 N and 1.59 mm and are in close agreement with the values presented in \cite{KruegerBenchmark} for the same DCB specimen. Therefore, the 0.25 mm mesh for the Abaqus CC model is considered converged and its solution taken as the reference for the subsequent model validation.

\subsection{Comparison with the proposed model}
\begin{figure}
	\centering
	\includegraphics[width=0.8\textwidth]{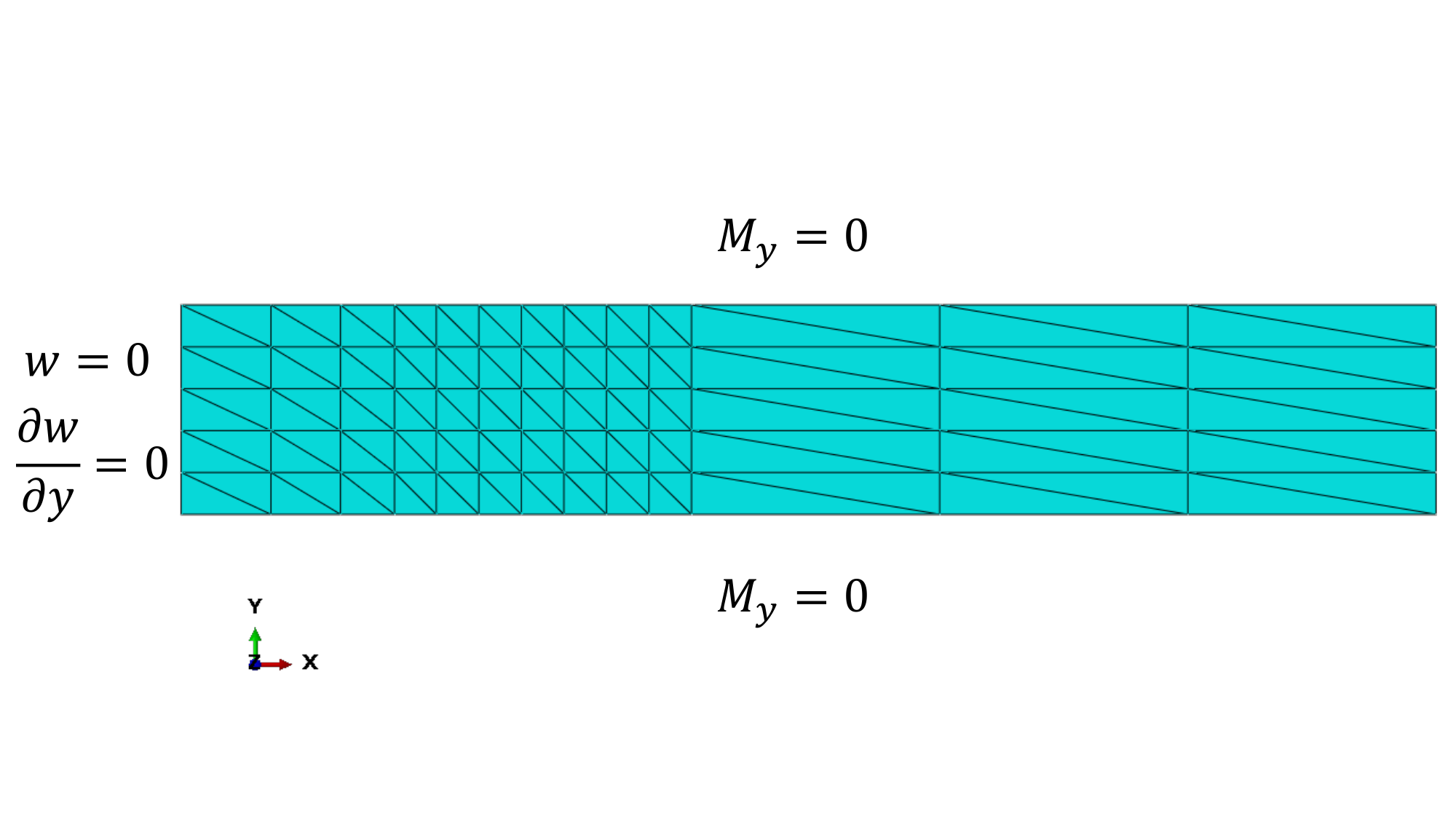}
	\caption{TUBA3 mesh and boundary conditions of the DCB bottom substrate.} 
	\label{fig:vl_tuba3_mesh_bcs}
\end{figure}
The TUBA3 models were generated following a different approach with respect to the Abaqus CC ones. First, a planar shell in a 3D space, corresponding to the bottom arm of the specimen, was created in Abaqus CAE and meshed for the required element size. Based on this model, a preliminary input file was produced. A Matlab program would then open the input file and write the nodal coordinates and connectivities for both the top arm elements and the CEs.
\par The scripts for computing the stiffness matrix and residuals vector for both TUBA3 and TUBA3-CE were included in a single user element subroutine, which could execute the code for the correct element based on a key passed by the Abaqus processor. The user element properties included the material data, thickness, and, in the case of TUBA3-CE, a binary flag variable, indicating whether the CE had been placed or not in the precracked region. In the former case, the pertaining damage variable was set equal to one. Otherwise, the default initial value for damage variable is equal to zero.
\par The boundary conditions are illustrated in \figref{fig:vl_tuba3_mesh_bcs}. Due to the use of curvature DOFs and the high order of interpolation of the TUBA3 elements, the enforcement of boundary conditions require special attention \cite{IVANNIKOV2015C1TUBAshell}. As the TUBA3 elements have curvature DOFs, the imposition of zero moment on edge nodes can be enforced explicitly. In the case of a free or simply supported edge, no moment exists around the edge axis. If the edge axis is oriented in the $x$ or $y$ direction and the plate material is isotropic, the followings must hold
\begin{align}
    M_y=\frac{E\,t^3}{12(1-\nu^2)}\Bigg(\nu\,\pdv[2]{w}{x}+\pdv[2]{w}{y}\Bigg)=0\\
    M_x=\frac{E\,t^3}{12(1-\nu^2)}\Bigg(\pdv[2]{w}{x}+\nu\,\pdv[2]{w}{y}\Bigg)=0
\end{align}
which reduce to
\begin{align}
    \nu\,\pdv[2]{w}{x}+\pdv[2]{w}{y}&=0\label{eq:AppC_My_iso}\\
    \pdv[2]{w}{x}+\nu\,\pdv[2]{w}{y}&=0\label{eq:AppC_Mx_iso}
\end{align}
Analogous conditions can be derived for composite laminates using the expressions of $M_y, M_x$ in terms of curvatures. As these conditions involve multiple curvature DOFs, they can be enforced through equation constraints.

\par As did in the reference Abaqus models, the analyses were carried out with the Full Newton-Raphson method, loaded on the top left edge of the specimen up to 4 mm displacement along z.

\subsubsection{Load-displacement curves}
\begin{figure}
	\centering
	\includegraphics[width=\textwidth]{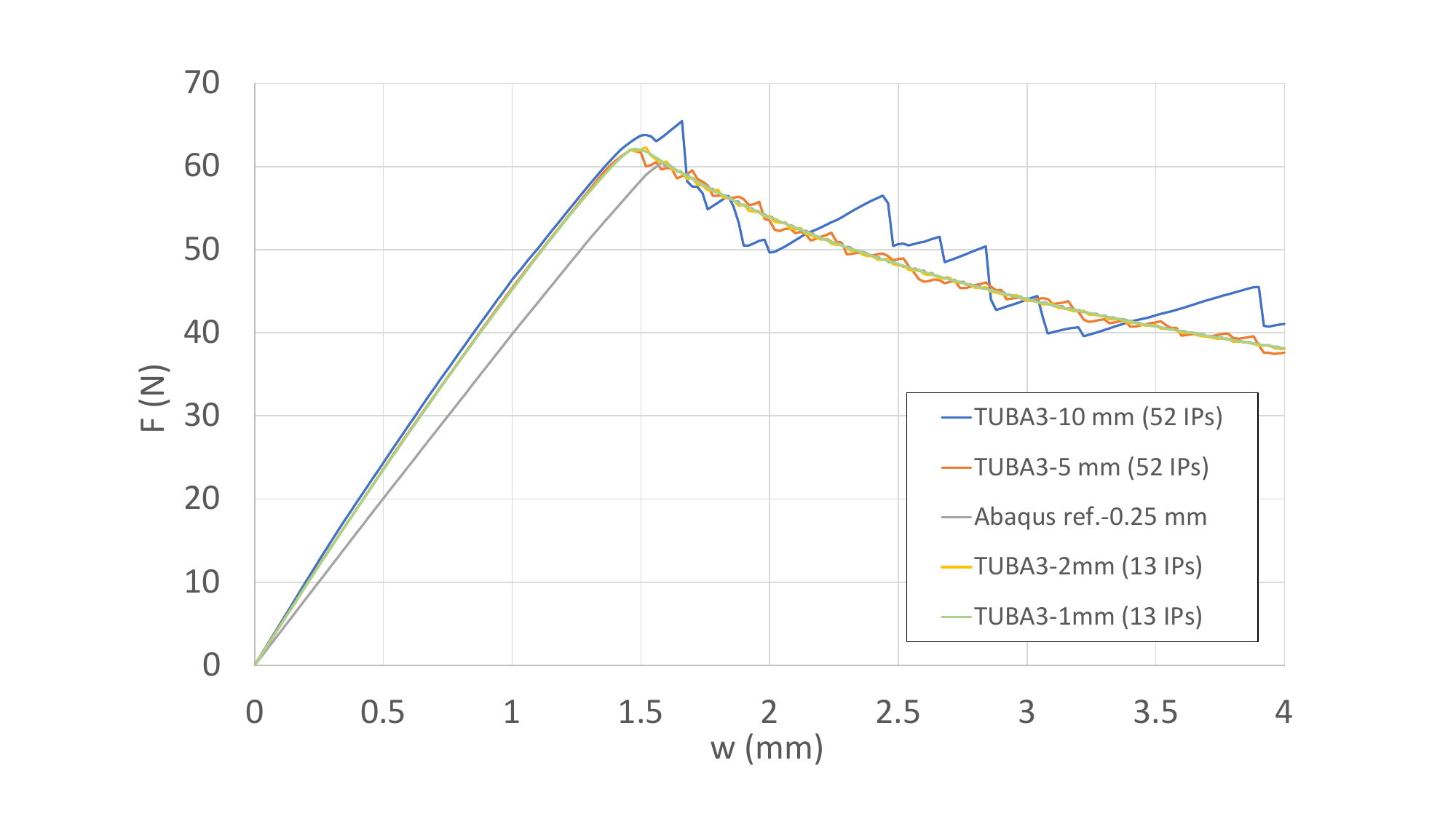}
	\caption{DCB load-displacement curves obtained with TUBA3 elements. Results for four different mesh sizes are plotted.} 
	\label{fig:vl_tuba3_loadisp}
\end{figure}
\figref{fig:vl_tuba3_loadisp} shows the DCB load-displacement curves obtained from the TUBA3 simulations. Results for four different element lengths are plotted together with the Abaqus reference solution. The simulations for mesh dimensions of 1, 2 and 5 mm achieved convergence and produced results in close agreement with the reference ones throughout the entire loading. Moreover, despite the quite severe oscillations during propagation, also the 10 mm mesh predicted the critical load and displacement with reasonable accuracy. As the cohesive zone is less than 1 mm in length as shown earlier in Figure~\ref{fig:stress_damage_2d}, this suggests an almost insensitivity of the TUBA3-CE to the cohesive zone.
\begin{table}[]
\centering
\begin{tabular}{@{}ccccccc@{}}
\toprule
\textbf{Model}       &  & $\bm{F_c}$ & \textbf{err\%} &  & $\bm{w_c}$ & \textbf{err\%} \\
                     &  & (N)           &                 &  & (mm)          &                 \\ \midrule
1 mm, 13 IPs         &  & 62.13         & 2.73            &  & 1.48          & 6.62            \\
2 mm, 13 IPs         &  & 62.31         & 3.02            &  & 1.52          & 4.10            \\
5 mm, 52 IPs         &  & 61.97         & 2.47            &  & 1.46          & 7.89            \\
10 mm, 52 IPs        &  & 63.78         & 5.47            &  & 1.52          & 4.10            \\
                     &  &               &                 &  &               &                 \\ \hline
\textbf{Abaqus ref.} &  & 60.48         &                 &  & 1.59          &                 \\ \bottomrule
\end{tabular}
\caption{Critical loads and displacements for TUBA3 DCB models of different mesh sizes. Abaqus reference solution and relative deviations are also reported.}
\label{tab:vl_critical_ld_err}
\end{table}
\par More quantitative measures of the accuracy reached with the new elements are given in Table \ref{tab:vl_critical_ld_err}. The 1, 2 and 5 mm meshes all managed to predict the critical load with at most 3\% error and the coarsest mesh model did not go over a 5.47\% deviation. Slightly worse precision was obtained on the critical displacements, whose error ranges from 4\% to almost 8\%. This discrepancy is linked to the difference in the initial stiffness between the TUBA3 and reference models, observed in \figref{fig:vl_tuba3_loadisp}. Note that TUBA3 elements only work in bending, whereas the C3D8I elements used for the reference can also deform out of plane. Therefore, the TUBA3 model would be stiffer than the C3D8I model. 
\begin{figure}
	\centering
	\includegraphics[width=\textwidth]{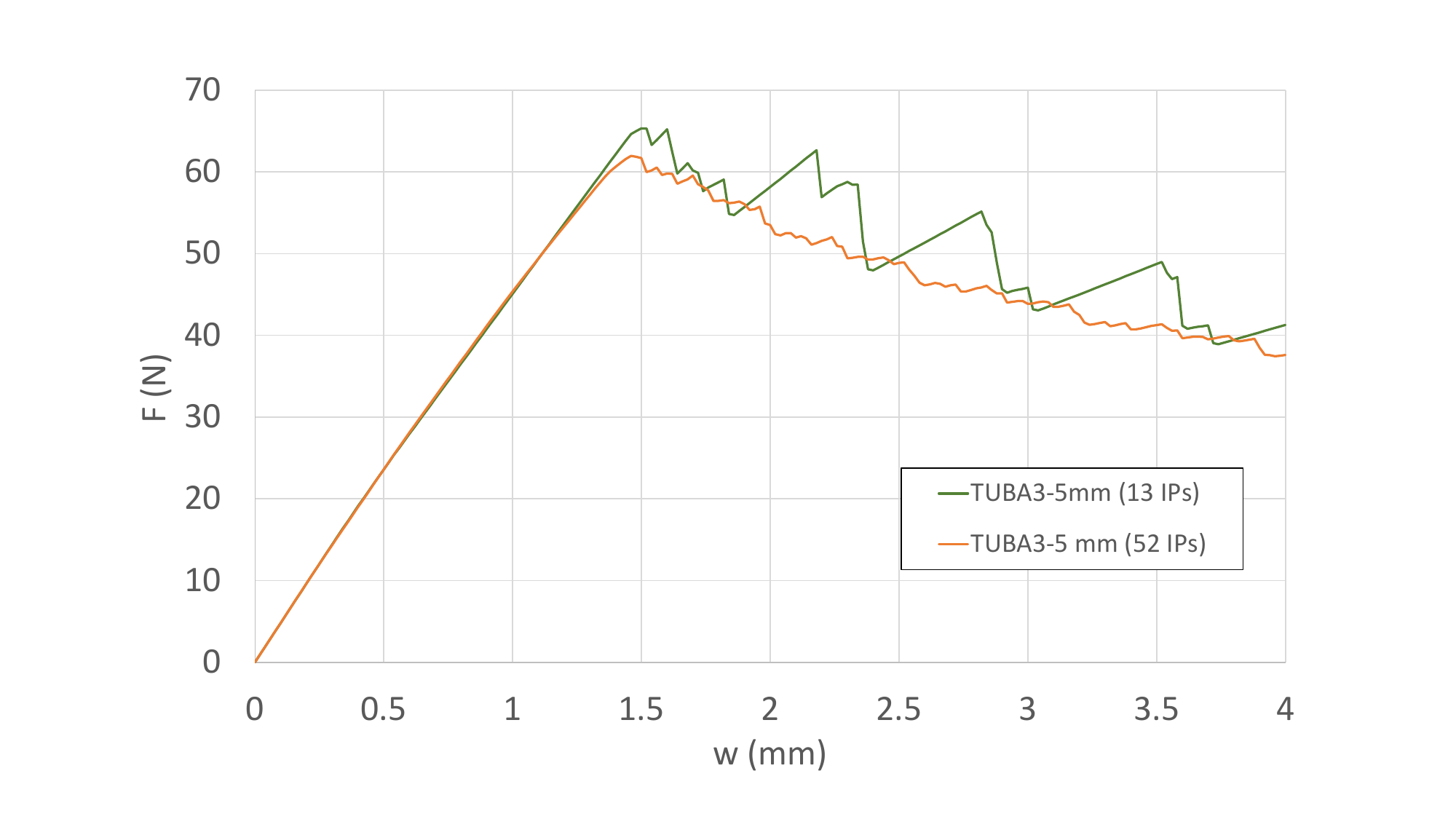}
	\caption{DCB load-displacement curves obtained with 5 mm TUBA3 elements. Curves for two different numbers of IPs are plotted.} 
	\label{fig:vl_tuba3_loadisp_IPs}
\end{figure}
\par A further point of attention concerns the number of IPs necessary to avoid instabilities or loss of convergence after damage. \figref{fig:vl_tuba3_loadisp_IPs} compares the load-displacement curves for two meshes of 5 mm element length in the fracture region. The TUBA3-CEs were given 13 IPs in one case and 52 in the other. The subdomain integration procedure described in Section \ref{ssec:TUBA3-CE_K} easily allowed to increase the number of IPs for each TUBA3-CE. Most noticeably, the curve corresponding to the lower IPs density results in spurious oscillations when the crack front is propagating. In fact, at every iteration of the analysis some of the IPs transition from an intact to a damaged state, removing part of the total stiffness. If few IPs are present, the failure of one of them causes a large stiffness loss, explaining the staggered profile in \figref{fig:vl_tuba3_loadisp_IPs}. Moreover, if the first line of IPs is too distant from the precrack front, these IPs would reach the failure opening $\Delta_I^f$ at a higher applied load than the case with more densely populated IPs. As observed in \figref{fig:vl_tuba3_loadisp_IPs}, the limit force is in fact higher in the 13 IPs case than when 52 IPs are used. In quantitative terms, the solution with 52 IPs overpredicts the critical load by 2.47\% its reference value, while the one with 13 IPs misses it by more than 8\%.
\par It could be argued that reducing the number of CEs at the cost of adding IPs would hinder the efficiency of the method. However, an increase in the number of elements ultimately enlarges the dimension of the FE system of equations, which would translate to a larger global stiffness matrix and more processing time to factorize it. On the other hand, using more IPs only results in a higher number of loops over the code portions that build the element stiffness matrix and residual vector. These loops could be parallelized as the IPs are independent from each other. A more quantitative discussion on the computational performances is offered in Section \ref{ssec:vl_CPU_perf}, which compares the CPU times of TUBA3 and reference models.

\subsubsection{Stress and damage profiles}
\begin{figure}
	\centering
	\includegraphics[width=\textwidth]{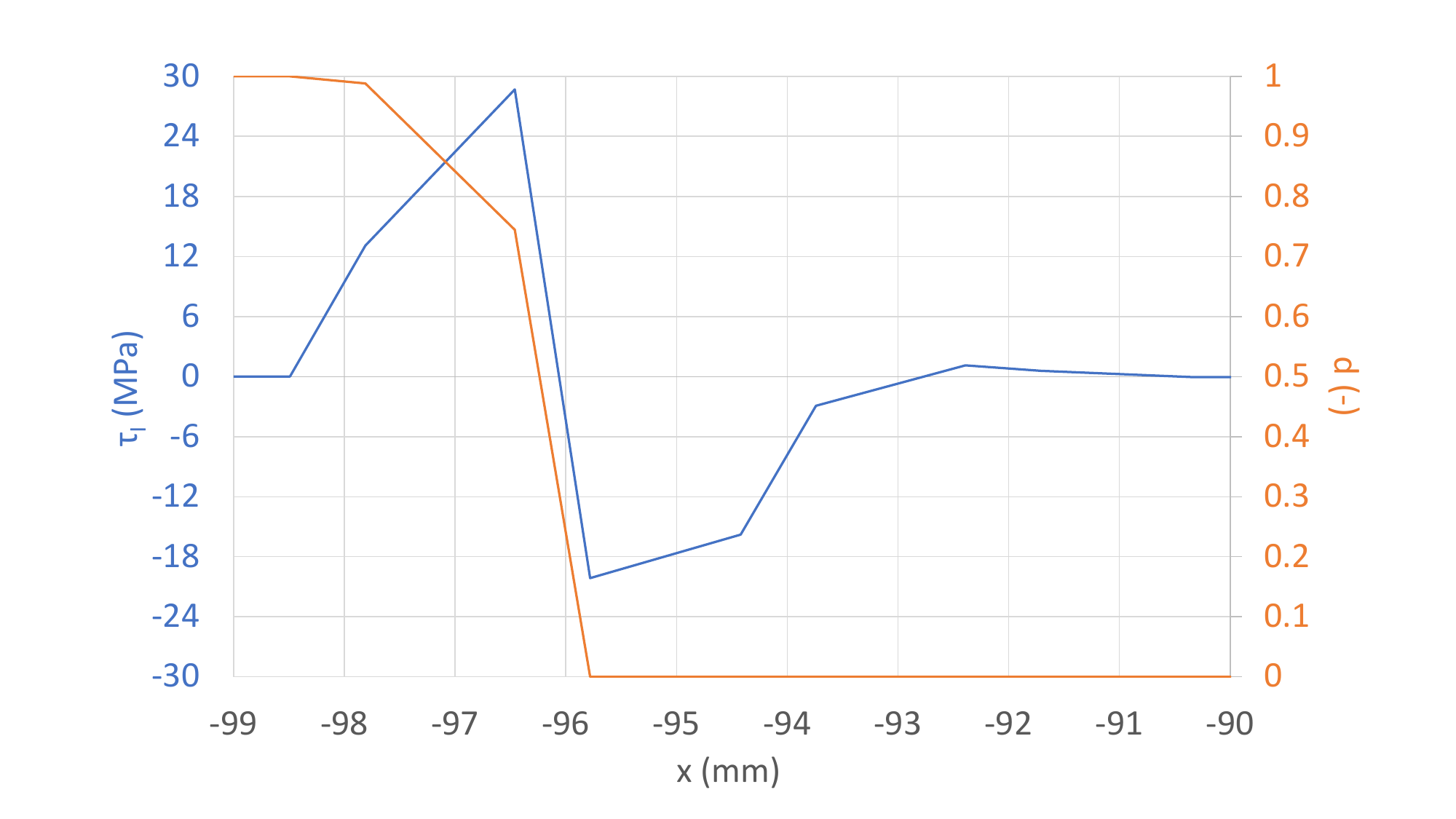}
	\caption{Stress and damage profiles in the cohesive zone of the DCB specimen. TUBA3 and TUBA3-CEs are used with a 2 mm size in the crack region and 13 IPs per element.}
	\label{fig:vl_tuba3_stress_damage_2mm}
\end{figure}

\begin{figure}
	\centering
	\includegraphics[width=\textwidth]{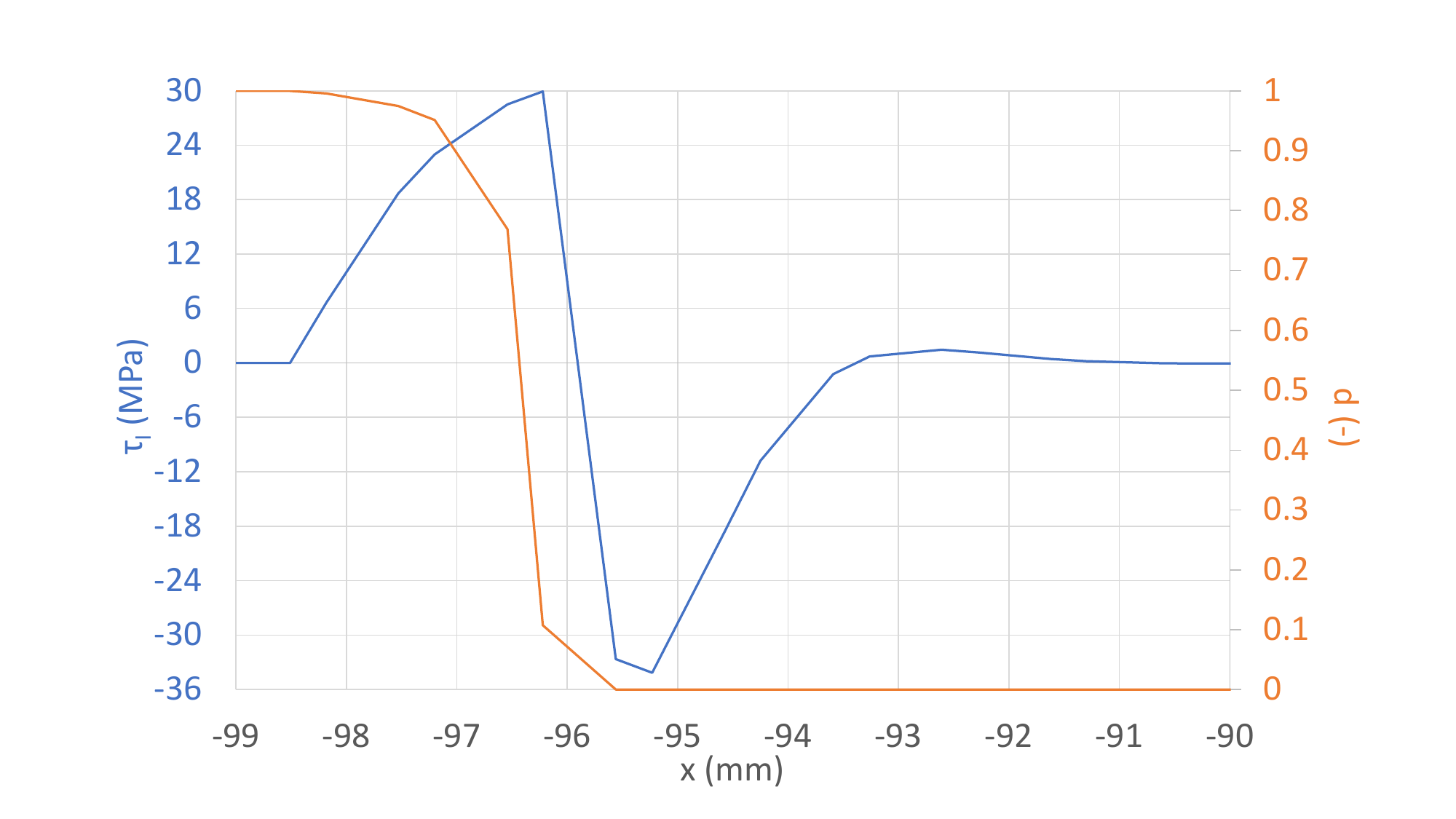}
	\caption{Stress and damage profiles in the cohesive zone of the DCB specimen. TUBA3 and TUBA3-CEs are used with 1 mm size in the crack region and 13 IPs per element.}
	\label{fig:vl_tuba3_stress_damage_1mm}
\end{figure}

The out-of-plane stresses and damage for the DCB specimen obtained with the 2 mm and 1 mm TUBA3 meshes are reported in \Cref{fig:vl_tuba3_stress_damage_2mm,fig:vl_tuba3_stress_damage_1mm}. Both these variables were computed for the CEs in the cohesive zone length at 4 mm opening of the specimen arms. The plots were produced with a Python program that would read the values of $\tau_I$ and $d$ at the barycentric integration point of a line of triangles. In particular the elements were chosen as equidistant from the specimen's longitudinal sides, in order to avoid edge effects.
\par Both graphs show the characteristic trends observed in the crack front region of a DCB specimen. From left to right, the interface is initially fully separated, hence no tractions are present and $d=$ 1. Proceeding further, the stresses increase and the damage variable decreases, identifying the beginning of the cohesive zone. The value of the material strength (30 MPa) is reached at the crack tip, where the material is intact. Immediately ahead of this region, negative stresses first arise to restore the interface equilibrium and then further assess to a zero plateau, where both arms of the specimen are still undeformed.
\begin{figure}
	\centering
	\includegraphics[width=\textwidth]{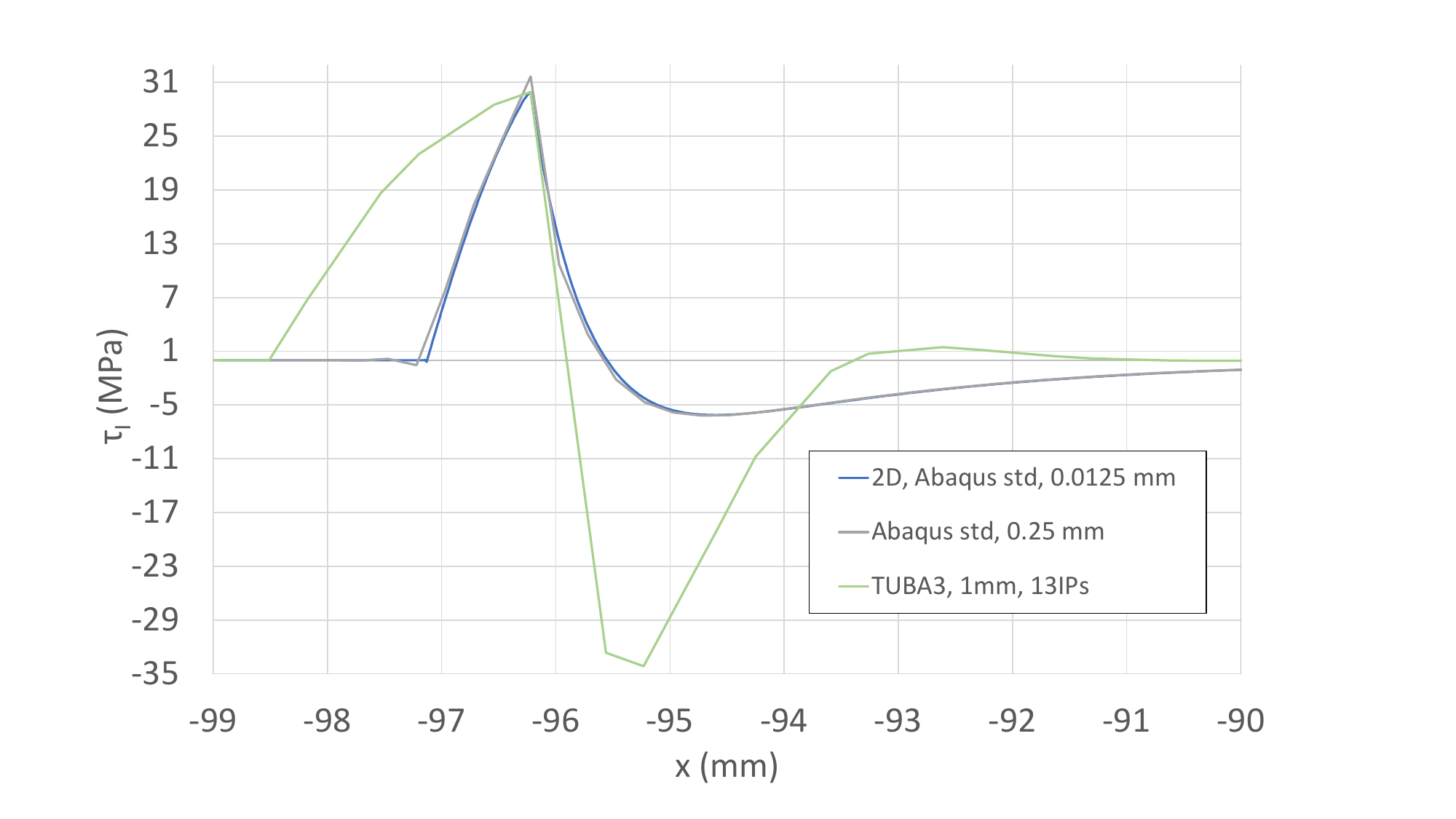}
	\caption{Comparison of stress profiles for different models and mesh sizes.}
	\label{fig:vl_comp_stresses}
\end{figure}
\begin{figure}
	\centering
	\includegraphics[width=\textwidth]{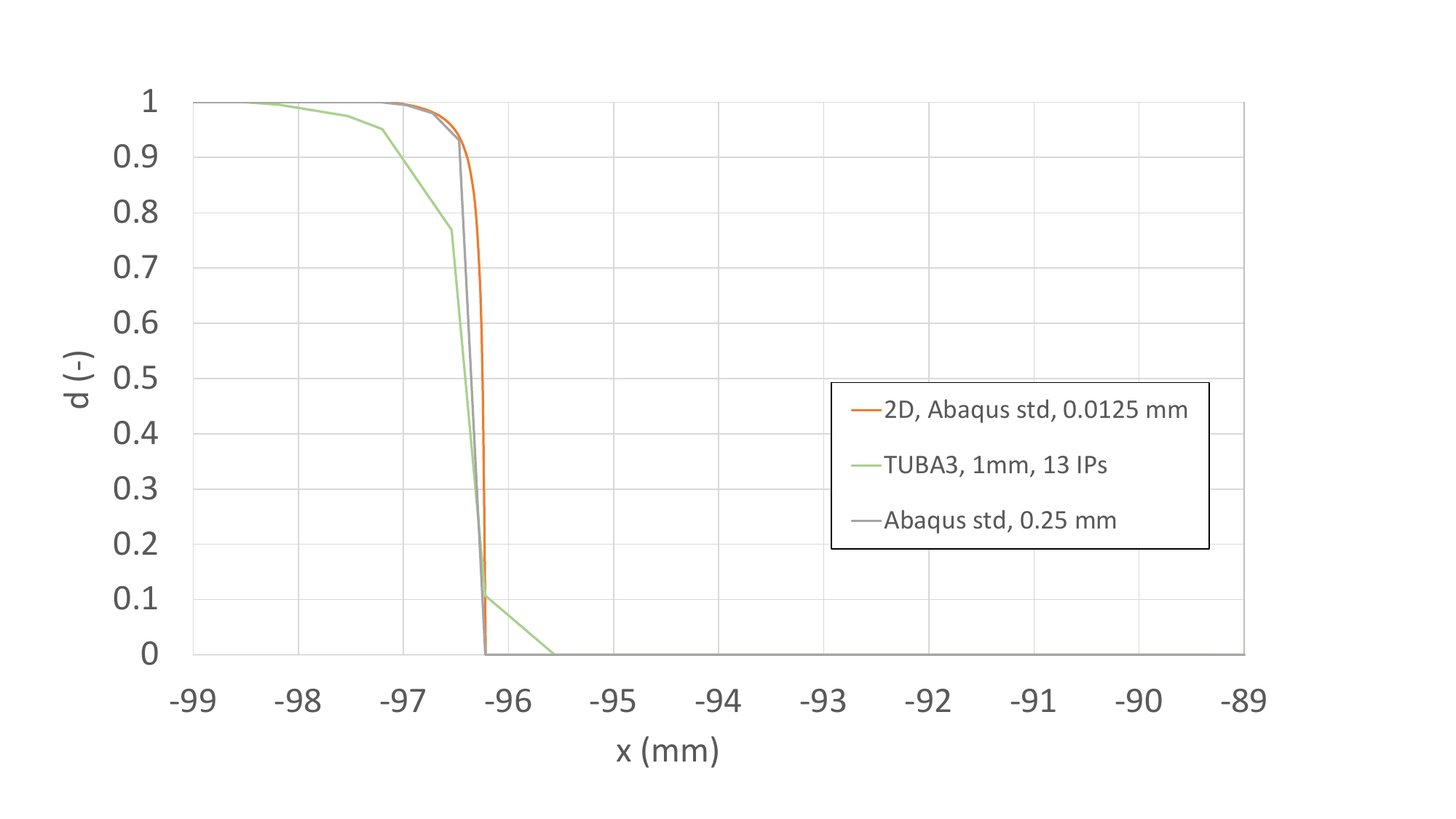}
	\caption{Comparison of damage profiles for different models and mesh sizes.}
	\label{fig:vl_comp_damage}
\end{figure}
\par \Cref{fig:vl_comp_stresses,fig:vl_comp_damage} compare stresses and damage in three different cases. Two of the curves plotted were obtained with 2D and 3D Abaqus CC models, respectively with 0.0125 and 0.25 mm element lengths, while the third one refers to the 1 mm TUBA3 model. Again, in order to avoid the edges influence, damage and stresses in Abaqus CC case were sampled in a line of nodes equidistant from the longitudinal sides of the specimen.
\par It is evident that the 0.25 mm Abaqus CC solution can be considered converged also in terms of stresses and damage, since the two fields match almost entirely the ones from the 2D analysis. The same does not hold for the TUBA3 models, which show visible deviations from the Abaqus CC results. As previously discussed, when plate elements are used as substrates, the out-of-plane stresses at the interface can only deform the CEs, which will open more than the case where they were included between solid elements. This explains why cohesive stress and damage start developing sooner in the TUBA3 model than in the CC models, when transitioning from left (the fully damaged region) to right (the intact region). 

\par Again looking at \figref{fig:vl_comp_stresses}, another relevant difference between Abaqus CC and TUBA3 models is the extent of the compression region and the stress magnitudes therein. Similarly as before, the discrepancy here can be attributed to the impossibility of plate elements to deform along their thickness. As soon as the crack is closed and the material is intact, the mid-planes of the plates come in contact and try to interpenetrate. This is almost entirely prevented by the large penalty stiffness, at the cost of generating high negative stresses. No alleviation of this effect comes from the compression and transverse shearing of the substrates along their thickness, as opposed to what happens with solid element models. 

\subsubsection{Damage maps}
\begin{figure}
	\centering
	\includegraphics[width=.8\textwidth]{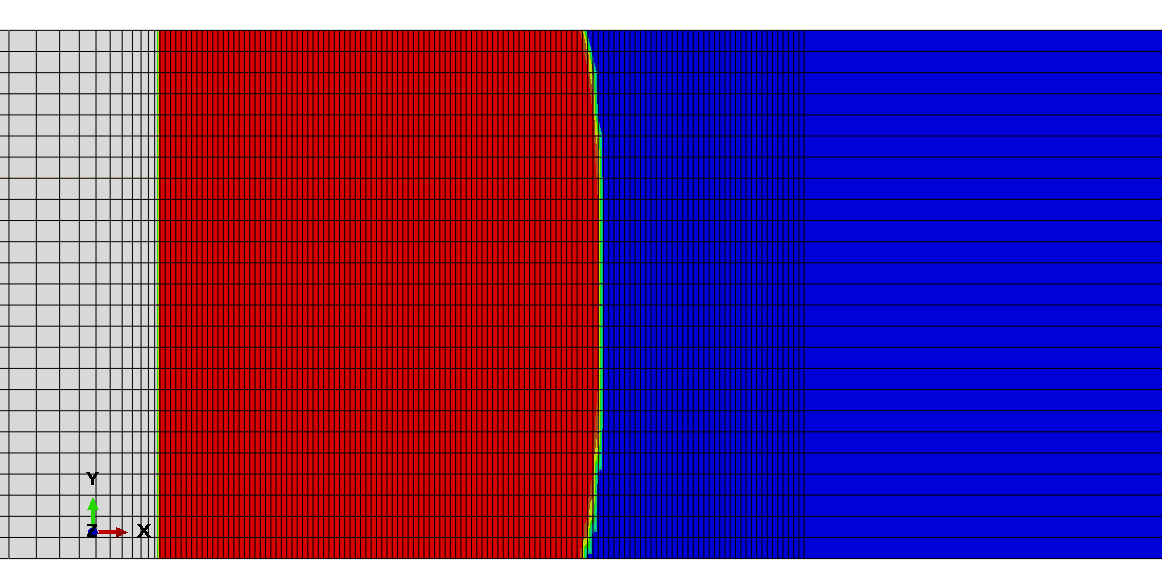}
	\caption{Damage distribution for the Abaqus CC model with 0.25 mm mesh at 4.0 mm opening of the DCB loaded edge; Red: fully damaged; Blue: intact; Green to orange: cohesive zone}
	\label{fig:cc_damage_map_zoom}
\end{figure}
The 2D interface damage distribution at final separation of the specimen's arms is shown in \figref{fig:cc_damage_map_zoom}. This plot was obtained from the Abaqus CC model with 0.25 mm mesh. Looking at the cohesive zone (green to orange coloured) it is clear how small this is with respect to the structural dimensions. It can be noticed how the crack front has the characteristic `thumbnail' shape. It is a well-known feature of pure mode I fracture observed both experimentally and numerically. During bending, the bonded surfaces of the substrates are both under tension, thus the Poisson's effect induces their compression in the direction orthogonal to the propagation one. This induces opposite curvatures for the two surfaces, closing the interface at the edges and delaying damage in these regions. 
\begin{figure}[p]
\centering
\includegraphics[width=0.9\textwidth]{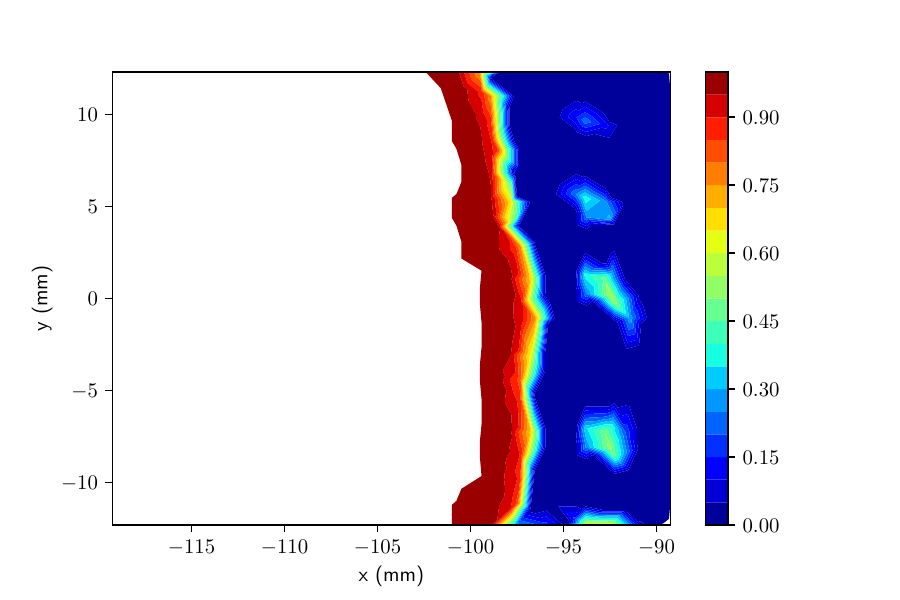}
\caption{TUBA3 model: damage distribution at 4.0 mm opening. Mesh size: 10 mm, 52 IPs.\label{fig:vl_damage_map-10mm-52IPs}}
\end{figure}
\begin{figure}[p]
\centering
\includegraphics[width=0.9\textwidth]{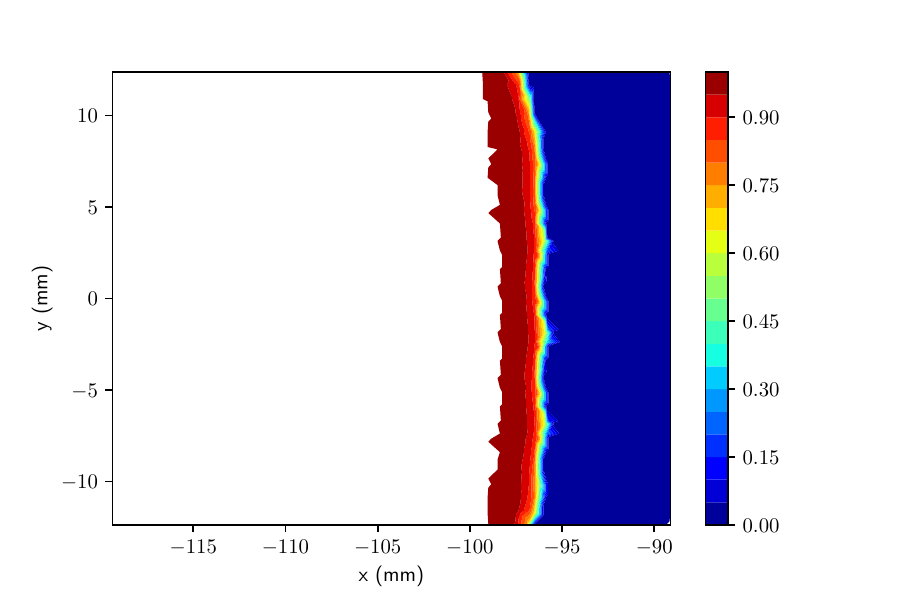}
\caption{TUBA3 model: damage distribution at 4.0 mm opening. Mesh size: 5 mm, 52 IPs.\label{fig:vl_damage_map-5mm-52IPs}}
\end{figure}
\begin{figure}[p]
\centering
\includegraphics[width=0.9\textwidth]{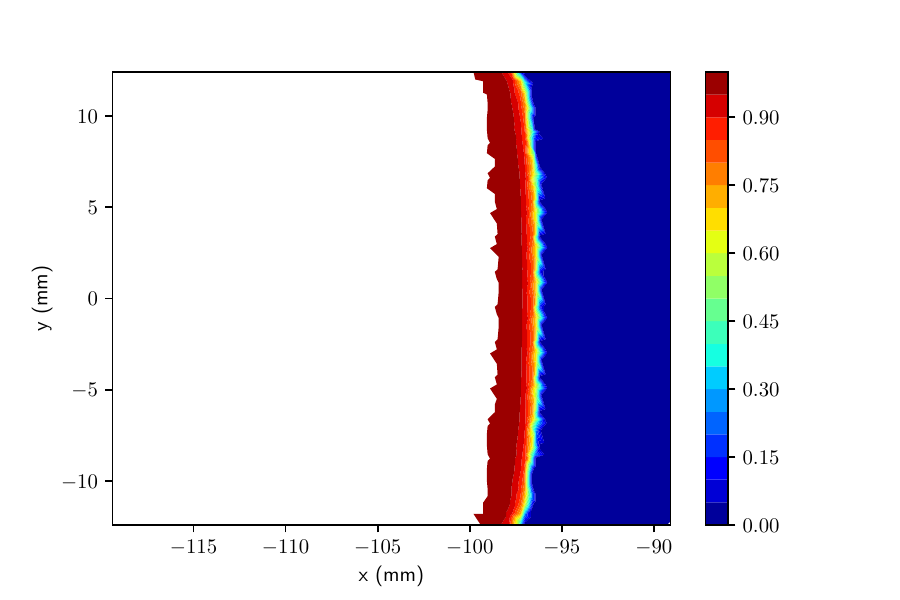}
\caption{TUBA3 model: damage distribution at 4.0 mm opening. Mesh size: 2 mm, 13 IPs.\label{fig:vl_damage_map-2mm-13IPs}}
\end{figure}
\begin{figure}[p]
\centering
\includegraphics[width=0.9\textwidth]{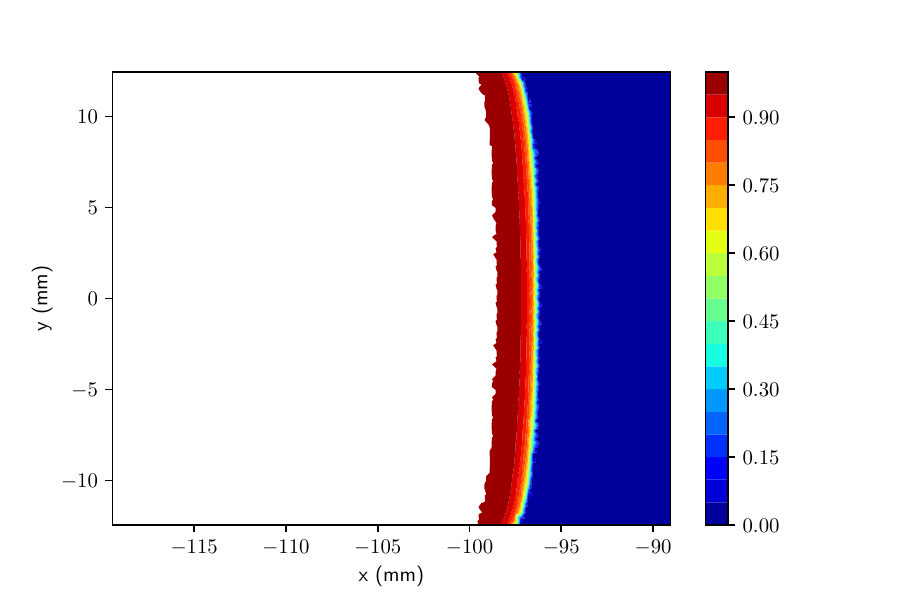}
\caption{TUBA3 model: damage distribution at 4.0 mm opening. Mesh size: 1 mm, 13 IPs.\label{fig:vl_damage_map-1mm-13IPs}}
\end{figure}
\par The damage maps for the TUBA3 models could also be produced by postprocessing the FE results with a Python program. The results are shown in Figures \ref{fig:vl_damage_map-10mm-52IPs} to \ref{fig:vl_damage_map-1mm-13IPs} for the same element sizes and number of IPs previously discussed. It is evident how finer meshes and higher number of IPs can smooth the damage distribution. The thumbnail shape becomes evident in the 1 mm and 2 mm cases and it can be compared with the reference in \figref{fig:cc_damage_map_zoom}. 
\par A numerical anomaly is noticed in the 10 mm case. The damage map for this mesh shows `damage spots' ahead of the crack front, where instead the material should be intact. A possible explanation is found observing the crack fronts for the meshes presented. All the damage maps show some `crests' in the variation of $d$ along $y$. The repetition of these crests seems periodic, with the period depending on the element's $y$-dimension. As previously discussed, the TUBA3 plate is allowed less deflection at its edges than in the rest of its domain (see Section \ref{ssec:tuba3_dofs_sfuns}). In the case of coarse meshes with large elements, this feature alters the stiffness of the structure depending on the directionality of the mesh. 

\subsubsection{Computational performances}
\label{ssec:vl_CPU_perf}
The final comparison between the TUBA3 models and the Abaqus CC solution concerns the performance parameters for the DCB simulations, reported in Table \ref{tab:vl_performance}. It is evident how even the most involved TUBA3 simulation manages to cut the CPU times of Abaqus CC of by an order of magnitude. Particularly impressive from a computational effort perspective is the saving obtained by running the 5 mm, 52 IPs simulation, still 97\% accurate on the limit load, as compared to the Abaqus CC, 0.25 mm case. The TUBA3 model achieves a 94.3\% reduction of the CPU time, an improvement similar to that scored in Ref.~\cite{RussoChen2019Jarticle} for the 2D case. 
\par All TUBA3 simulations show an higher iterations count than the Abaqus CC one. This difference makes sense considering that viscous regularization is used only for the Abaqus CC analyses. However if viscous regularization is removed,  Abaqus CC simulations could not converge, where the analysis would stop before the specimen reaches the 4 mm opening. On the contrary, TUBA3 models do not require viscous regularization to reach convergence, indicating better numerical stability of the TUBA3-CEs. 

\begin{table}[h]
\centering
\begin{tabular}{@{}ccccccc@{}}
\toprule
                        & \textbf{Abaqus CC} & \textbf{TUBA3} & \textbf{TUBA3} & \textbf{TUBA3} & \textbf{TUBA3} &\\ Element size & 0.25 mm & 1 mm & 2 mm & 5 mm & 10 mm &  \\ No. IPs & 4 & 13 & 13 & 52 & 52 &  \\ \midrule
\textbf{CPU time (s)}       & 12582                     & 1577.9                     & 1470.6                     & 718.24                     & 216.05                      &  \\
\textbf{No. DoFs}       & 181728                      & 14664                        & 4200                         & 1656                         & 480                           &  \\
\textbf{No. elements}   & 21496                       & 6900                         & 1872                         & 660                          & 144                           &  \\
\textbf{No. iterations} & 2040                        & 2249                         & 7722                         & 11091                        & 5522                          &  \\ \bottomrule
\end{tabular}
\caption{Computational performance parameters of the DCB specimen simulations for both Abaqus CC and TUBA3 models.}
\label{tab:vl_performance}
\end{table}

\section{Conclusions}
\label{sec:CFW}
This research explored the use of $C^1$ thin plate and compatible CEs as a novel approach to model 3D delamination, with the objective of tackling the most limiting problem of CEs, namely the cohesive zone limit on mesh density under Mode I fracture between thin substrates. The TUBA3 triangular thin plate element has been adopted for the modelling of the thin substrates and the TUBA3-CE has been developed here to model their delamination.

The classical composite DCB problem was used to assess the performance of the proposed TUBA3-based method against that of the standard CE approach. The load displacement curves of the TUBA3 models showed errors below 6\% of the limit load, using elements 11 times larger than the CZL. This led to a 94\% CPU time saving over the standard CE models. However,  due to the kinematic assumptions made by the classical plate theory, the out of plane straining of the substrates could not be reproduced by the proposed method. This led to inaccuracies in the predictions of the local stress and damage profiles. In this work, the global structural responses and limits are of the most interest, hence this deficiency is not deemed critical as it does not show impact on the global structural predictions. However, if the accurate predictions of local fields near the cohesive zone are also of interest, further development would be needed to improve or enrich the local predictions of the model. The use of curvature DOFs and the high order of interpolation require special care when imposing boundary conditions, making their practical engineering applications complicated. Ongoing work in the research group is exploring Kirchhoff plate element formulations without curvature DOFs to replace the role of TUBA3 element here, while retaining the demonstrated capability of overcoming the cohesive zone limit.

\bibliographystyle{elsarticle-num-names}
\bibliography{MyBib}

\appendix

\section{TUBA3 shape functions}
\label{App:ApA}
A planar triangle is considered, of vertices $(x_1,y_1)$, $(x_2,y_2)$ and $(x_3,y_3)$. The following quantities can be defined
\begin{equation}
    \begin{cases}
        a_i=&x_jy_k-x_ky_j\\
        b_i=&y_j-y_k\\
        c_i=&x_k-x_j
    \end{cases}
    \label{eq:AppE_abc}
\end{equation}
\begin{equation}
    r_{ij}=-\frac{b_ib_j+c_ic_j}{b_i^2+c_i^2}
    \label{eq:AppE_r}
\end{equation}
where $i$,$j$,$k$ are cyclic permutations of 1,2,3.
\par Using the definitions in \Cref{eq:AppE_abc,eq:AppE_r}, the first six shape functions of the TUBA3 triangle, as reported in \cite{Dasgupta1990ARevisited}, are
\allowdisplaybreaks
\begin{align*}
    N_1(L_1, L_2, L_3) = &L_1^5 + 5 L_1^4 L_2 + 5 L_1^4 L_3 + 10 L_1^3 L_2^2 + 10 L_1^3 L_3^2 + 20 L_1^3 L_2 L_3\\&+ 30 r_{21} L_1^2 L_2 L_3^2 + 30 r_{31} L_1^2 L_2 ^2 L_3\\
    N_2(L_1, L_2, L_3) = &c_3 L_1^4 L_2 - c_2 L_1^4 L_3 + 4 c_3 L_1^3 L_2^2 - 4 c_2 L_1^3 L_3^2 + 4 (c_3 - c_2) L_1^3 L_2 L_3\\&- (3 c_1 + 15 r_21 c_2) L_1^2 L_2 L_3^2 + (3 c_1 + 15 r_{31} c_3) L_1^2 L_2^2 L_3\\
    N_3(L_1, L_2, L_3) = &-b_3 L_1^4 L_2 + b_2 L_1^4 L_3 - 4 b_3 L_1^3 L_2^2 + 4 b_2 L_1^3 L_3^2 + 4 (b_2 - b_3) L_1^3 L_2 L_3\\&+ (3 b_1 + 15 r_{21} b_2) L_1^2 L_2 L_3^2 - (3 b_1 + 15 r_{31} b_3) L_1^2 L_2^2 L_3\\
    N_4(L_1, L_2, L_3) = &\frac{c_3^2}{2} L_1^3 L_2^2 + \frac{c_2^2}{2} L_1^3 L_3^2 - c_2 c_3 L_1^3 L_2 L_3 +\Bigg(c_1 c_2 + \frac{5}{2} r_{21} c_2^2\Bigg) L_1^2 L_2 L_3^2 \\&+\Bigg(c_1 c_3 + \frac{5}{2} r_{31} c_3^2\Bigg) L_1^2 L_2^2 L_3\\
    N_5(L_1, L_2, L_3) = &-b_3 c_3 L_1^3 L_2^2 - b_2 c_2 L_1^3 L_3^2 + (b_2 c_3 + b_3 c_2) L_1^3 L_2 L_3\\&- (b_1 c_2 + b_2 c_1 + 5 r_{21} b_2 c_2) L_1^2 L_2 L_3^2 - (b_1 c_3 + b_3 c_1 + 5 r_{31} b_3 c_3) L_1^2 L_2^2 L_3\displaybreak[3]\\
    N_6(L_1, L_2, L_3) = &\frac{b_3^2}{2} L_1^3 L_2^2 +\frac{b_2^2}{2} L_1^3 L_3^2 -b_2 b_3 L_1^3 L_2 L_3 + \Bigg(b_1 b_2 + \frac{5}{2} r_{21} b_2^2\Bigg) L_1^2 L_2 L_3^2\\&+ \Bigg(b_1 b_3 + \frac{5}{2} r_{31} b_3^2\Bigg) L_1^2 L_2^2 L_3\\
\end{align*}
where $L_1$,$L_2$ and $L_3$ are the triangle's area coordinates.
\par The remaining shape functions ($N_7$ to $N_{18}$) are defined in sets of six, by cyclically permuting the indices $1$,$2$ and $3$.







\end{document}